\documentclass[sigconf, authorversion, nonacm]{acmart}

\AtBeginDocument{%
  }

\setcopyright{acmlicensed}
\copyrightyear{2018}
\acmYear{2018}
\acmDOI{XXXXXXX.XXXXXXX}

\acmConference[Conference acronym 'XX]{Make sure to enter the correct
  conference title from your rights confirmation emai}{June 03--05,
  2018}{Woodstock, NY}

\acmISBN{978-1-4503-XXXX-X/18/06}
\begin{document}

\title{Synthetic Human Memories: AI-Edited Images and Videos Can Implant False Memories and Distort Recollection}

\author{Pat Pataranutaporn}
\affiliation{%
  \institution{MIT Media Lab, Massachusetts Institute of Technology}
  \city{Boston}
  \state{Massachusetts}
  \country{USA}
}
\email{patpat@media.mit.edu}

\author{Chayapatr Archiwaranguprok}
\affiliation{%
  \institution{University of the Thai Chamber of Commerce}
  \city{Bangkok}
  \country{Thailand}
}
\email{chayapatr.arc@gmail.com}

\author{Samantha W. T. Chan}
\affiliation{%
  \institution{MIT Media Lab, Massachusetts Institute of Technology}
  \city{Cambridge}
  \state{Massachusetts}
  \country{USA}
}
\email{swtchan@media.mit.edu}

\author{Elizabeth Loftus}
\affiliation{%
  \institution{UCI School of Social Ecology, UC Irvine}
  \city{Irvine}
  \state{California}
  \country{USA}
}
\email{eloftus@law.uci.edu}

\author{Pattie Maes}
\affiliation{%
  \institution{MIT Media Lab, Massachusetts Institute of Technology}
  \city{Cambridge}
  \state{Massachusetts}
  \country{USA}
}
\email{pattie@media.mit.edu}

\renewcommand{\shortauthors}{Pataranutaporn et al.}

\begin{abstract}


AI is increasingly used to enhance images and videos, both intentionally and unintentionally. As AI editing tools become more integrated into smartphones, users can modify or animate photos into realistic videos. This study examines the impact of AI-altered visuals on false memories—recollections of events that didn’t occur or deviate from reality. In a pre-registered study, 200 participants were divided into four conditions of 50 each. Participants viewed original images, completed a filler task, then saw stimuli corresponding to their assigned condition: unedited images, AI-edited images, AI-generated videos, or AI-generated videos of AI-edited images. AI-edited visuals significantly increased false recollections, with AI-generated videos of AI-edited images having the strongest effect (2.05x compared to control). Confidence in false memories was also highest for this condition (1.19x compared to control). We discuss potential applications in HCI, such as therapeutic memory reframing, and challenges in ethical, legal, political, and societal domains.

\end{abstract}



\begin{teaserfigure}
  \includegraphics[width=1\textwidth]{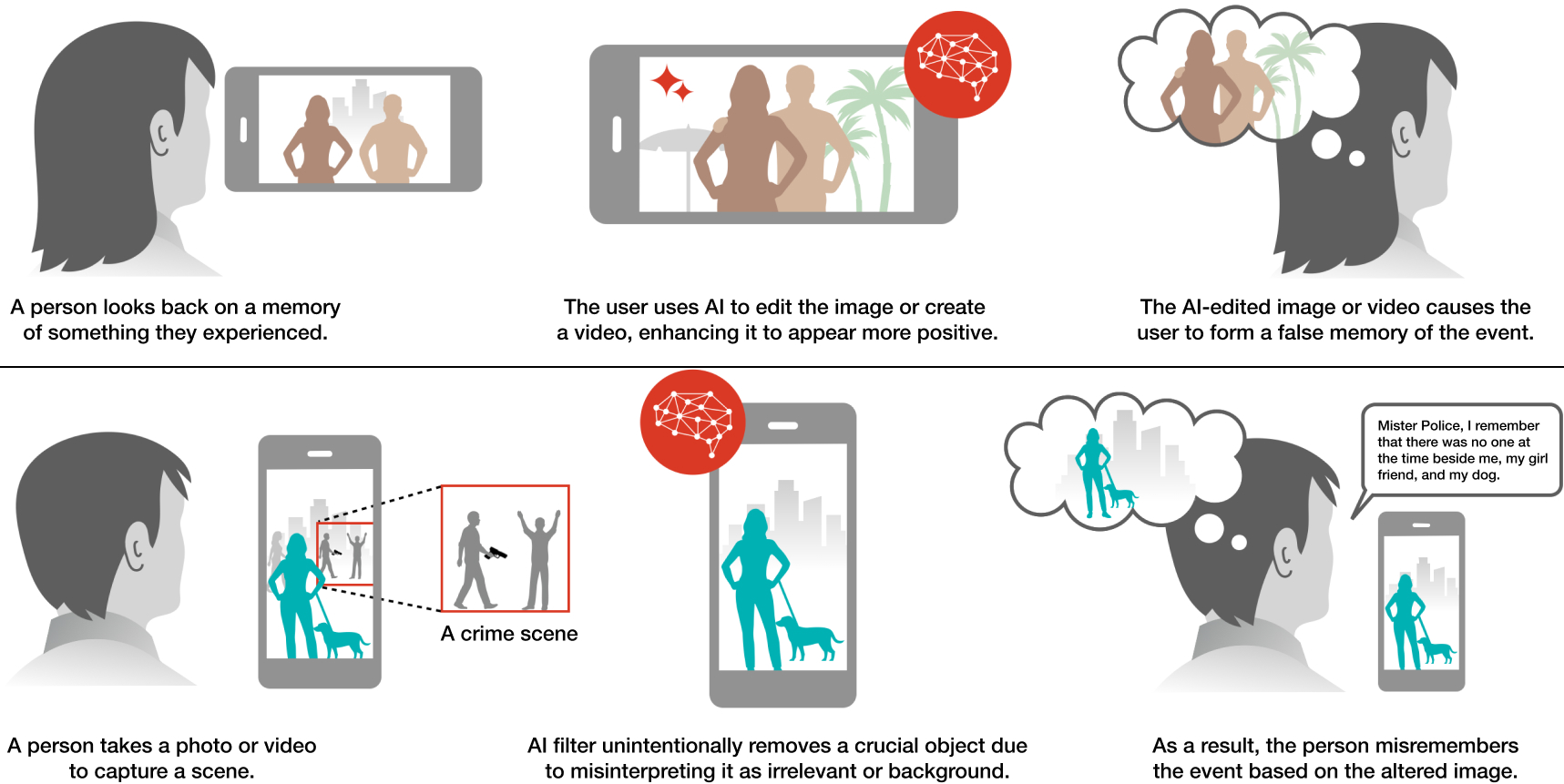}
  \caption{Illustration of how AI-edited media can create false memories. The top row depicts a person using AI to enhance an image or video to make it more positive. Over time, the person revisits the image without recalling that it was edited, leading to the development of a false memory of the event. The lower section depicts a situation where AI inadvertently modifies an image, eliminating bystanders from the frame as part of an automatic filter without retaining the original version (a feature already available in Google Photos and other camera apps). Later, when the individual reviews the photograph—potentially related to a crime scene—they develop a false recollection that matches the edited image rather than the actual event, leading to false witness testimony. This figure highlights the impact of AI-generated edits on human memories, demonstrating how subtle changes can distort recollection.}
  \label{fig:teaser}
\end{teaserfigure}


\maketitle

\section*{Introduction}
If a device existed that could help reframe your worst day in a more positive light, would you choose to use it? Memory-editing technologies have been a central theme in science fiction, prominently featured in works such as Eternal Sunshine of the Spotless Mind, Men in Black, Total Recall, and Inception \cite{phelps2019memory}. However, techniques for altering human memories are not confined to the realm of fiction, as they represent a heavily studied area within psychology and cognitive science ~\cite{Muschalla2021}.

False memories, which refer to recollections of events that either never occurred or are significantly distorted from reality, have been a major focus in psychology research. The study of false memories is vital because they can distort witness testimonies, disrupt legal processes, and lead to faulty decision-making based on incorrect information. Given these broad implications, understanding how false memories form is a critical area of investigation \cite{loftus2003make, slotnick2004sensory, gonsalves2000neural, zhuang2022rapid, loftus1997creating, loftus1995formation}. Unlike typical forms of misinformation \cite{zhou2023synthetic}, false memories are particularly insidious because the individual genuinely believes they recall accurate events, making them resistant to correction and potentially more influential in shaping beliefs and behaviors~\cite{Loftus1992, Loftus2005}. Moreover, false memories can serve as a seed for making people more susceptible to additional false information~\cite{Thorson2015, Kan2021}, creating a cascading effect that further distorts perceptions of reality and complicates efforts to establish accurate historical or personal narratives.

Research by Loftus and Palmer \cite{loftus1974reconstruction} showed how the wording of questions can significantly influence eyewitness memory. When participants watched a video of a car accident, their estimates of speed estimates varied based on the verb used in the question (e.g., “collided,” “smashed”, “bumped,” “contacted,” or “hit”). Additionally, the “Lost in the Mall” study \cite{loftus1995formation} demonstrated that entirely false childhood memories could be implanted. In a more recent replication \cite{murphy2023lost}, a larger sample size revealed that 35\% of participants reported a false memory of being lost in a mall during childhood, compared to 25\% in the original experiment. These findings strengthen the validity of the initial study and emphasize the relevance of such phenomena in legal settings.

In the context of visually induced false memories \cite{stephan2017visual, robin2022effects, wang2018nature}, researchers have shown that exposure to images of fictitious events, such as a hot-air-balloon ride, can lead to the formation of false memories related to the depicted experience \cite{garry2005photographs}. Researchers have employed various methods to visually induce false memories, including single presentations of thematic scenes with omitted elements \cite{miller1998creating}, personal photographs \cite{lindsay2004true}, and narrative instructions during interviews \cite{wade2002picture}. In a notable study, fifty percent of participants developed complete or partial false memories after exposure to a fake childhood photograph and guided imagery exercises over three interviews. These findings have significant implications for clinical and legal professionals working with memory-related cases \cite{wade2002picture}.

However, these studies have predominantly been conducted in controlled laboratory settings, where images are manually edited by researchers and interviews are carefully planned. The process also involves human intervention in establishing trust, guiding participants, and presenting the manipulated images, which inherently limits the scope and scale of false memory induction. With recent advancements in artificial intelligence (AI), however, these limitations are beginning to change. AI has the potential to automate and scale the creation and presentation of manipulated content, significantly expanding the possible impact of false memories on individuals.


AI-powered image enhancement is also increasingly becoming a standard feature in smartphones and cameras, operating both at capture and during post-processing. At the moment of capturing, AI may automatically remove unwanted elements or combine the best parts from multiple shots, as seen in Google's "Best Take" feature~\cite{besttake}. After the shot, applications such as Apple Photos, Google Photos, and Samsung's Galaxy AI provide AI tools by default for users to edit, remove, or alter photo features with just a few steps. Additionally, generative AI models, such as OpenAI’s Sora, Luma’s Dream Machine, and Kling, are increasingly used to animate static images into realistic-looking videos.

\begin{figure*}
    \centering
    \includegraphics[width=1\linewidth]{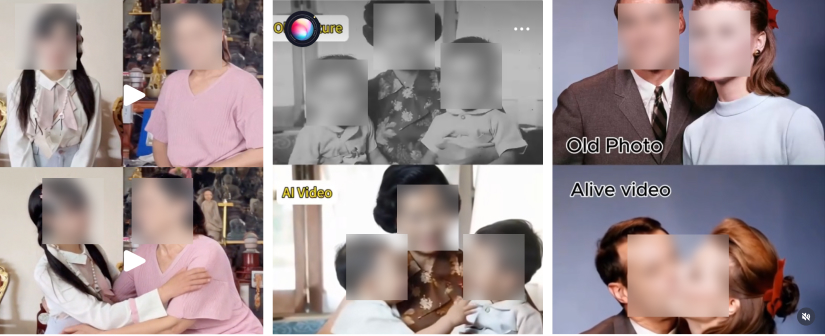}
    \caption{The figure illustrates how AI-generated content can potentially create false memories, particularly through AI-altered videos on social media platforms like TikTok. A recent trend on these platforms involves using AI to animate photos of deceased relatives, creating simulated interactions. These artificial experiences may blur the line between genuine memories and digitally fabricated ones, potentially affecting how people remember their loved ones.}
    \label{fig:false-memory-example}
\end{figure*}

The unprecedented proliferation of AI-driven image editing and video manipulation technologies has raised significant concerns regarding the integrity of consumed information. We argue that AI-generated content contributes to misinformation by distorting our understanding of the present (e.g., deepfakes) as well as reshaping how we remember the past. AI-generated media can potentially create false memories and lead individuals to recall past events differently than they actually occurred and were initially experienced. The implications of these technologies span both personal and societal domains, as illustrated in figure~\ref{fig:false-memory-example}.

On a personal level, there has been a notable trend, particularly on social media platforms such as TikTok, of users employing AI to animate photographs of deceased family members, simulating interactions with departed loved ones. On a broader scale, the potential for AI-generated content to influence collective memory and historical narratives poses significant challenges to societal understanding and cohesion, potentially altering public perceptions of past events and shaping future decision-making processes. For example, AI-edited images of public gatherings or protests could subtly alter the perceived scale or mood of these events, gradually reshaping how participants and observers remember their personal experiences and consequently influencing the collective memory of significant social movements.

A crucial distinction must be made between deepfakes and AI-edits, as both leverage generative AI but differ significantly in their real-world implications and how people encounter them. Deepfakes typically involve the creation of entirely fabricated audio or video content, often for malicious purposes such as spreading disinformation. In contrast, AI-edits modify existing content, subtly altering genuine memories or experiences. This distinction is important, as people may be more vigilant against obviously fake content but less aware of slight modifications to their own experiences or memories. While initial research has primarily explored the effects of deep fakes on false memory formation in political contexts~\cite{Liv2020, Murphy2021, tas2023changing}, this study specifically focuses on exploring the effects and implications of AI-edited content.

While AI-edited content poses risks for distorting memories, it also holds potential for positive applications for human memory. AI-assisted memory modification could help individuals process traumatic experiences more effectively, potentially reducing symptoms of conditions like post-traumatic stress disorder (PTSD) or depression. In addition, AI could assist in the creation of more comforting or positive recollections by emphasizing moments of joy or success. For example, AI might be used to enhance a photo from a difficult period by adjusting the atmosphere—making a gloomy day look sunny or adding vibrant details to a faded memory. 

The use of AI in memory alteration raises significant ethical questions and requires careful consideration of potential long-term psychological effects. The balance between therapeutic benefit and the integrity of personal memories must be carefully weighed in any application of such technology.

Motivated by both the potential risks and promising applications of AI-edited content in memory formation, this study investigates the effects of AI-edited images and videos on the formation of false memories. Specifically, we aim to assess whether exposure to AI-edited visuals can influence individuals’ recollections of past experiences. Particularly, this paper explores the following research questions:
\begin{enumerate}
    \item To what extent do AI-edited images influence participants’ memories of the original scenario, and how does this compare to participants in a control condition who view unedited images?
    \item Does the conversion of images into AI-generated videos further exacerbate the formation of false memories, both in terms of the number of false memories reported and the confidence participants have in these memories, compared to static AI-edited images and the control condition?
    \item How do different types of AI editing (e.g., changes to objects, people, or the environment) affect the severity of false memories?
    \item What factors, such as familiarity with AI filters, memory efficacy, skepticism, age, gender, and education, moderate the severity and frequency of false memory formation when participants are exposed to AI-edited media?
\end{enumerate}

In this pre-registered experiment involving 200 participants, subjects were initially shown a set of 24 images designed to serve as base memories. After completing a filler task, participants were then presented with a second set of visuals that had been modified by AI based on the experimental condition to which they were randomly assigned. These modifications included changes such as adding, removing, or altering objects, people, or environmental elements, thereby shifting the original context and meaning of the images. For example, some alterations included increasing the military presence in a scene, removing visible effects of climate change, or changing the personal attributes of individuals depicted. In video conditions, the images (edited or unedited) are transformed into videos using image-to-video AI, further enhancing the dynamism and altering the interpretation of the original visuals.

After viewing the unedited set and doing a filler task, participants were divided into four groups: one group viewed the original, unedited images (control), another viewed AI-edited images, a third group viewed AI-generated videos of the unedited images, and the final group viewed AI-generated videos of the AI-edited images. 


Our results demonstrate that AI-edited images and videos led to a significant increase in false recollections of the original scenario, with participants recalling incorrect details when exposed to the altered visuals, compared to the control condition. Notably, AI-generated videos of AI-edited images had the most profound effect on the number of reported false memories (2.05x compared to control). The confidence in false memories was also higher for AI-generated videos of AI-edited images (1.19x compared to control). Our study included a weighted score analysis and group homogeneity checks to validate the experimental design. 

Additional analyses exploring the effects across different types of image content (daily life, news, documentary) and elements edited by AI (people, objects, environments) consistently showed increased false memories with AI manipulation, with people-related edits having the highest absolute number of false memories (AI-generated videos of AI-edited image condition induced 45.3\% false memories of all memories reported) and environmental-related edits gaining the most dramatic increase (2.2x compared to control). A mixed-effects regression model identified age as a small but significant factor in false memory formation, while other demographic and cognitive factors showed no significant relationship.

To our knowledge, this is one of the first studies with empirical evidence demonstrating the impact of generative AI-based visual editing on the formation of false memories. Furthermore, we believe this is the first study to examine the impact of manipulated videos on the formation of false memories.

The paper also examines ethical considerations related to AI use in media and communication, including its potential influence on personal and collective memories. We explore the implications of these findings and discuss possible approaches to address AI-generated misinformation. The paper also touches on the societal, political, and legal aspects of memory alteration through AI technology.

Our research indicates that “externalizing” human memories, a concept long explored in Human-Computer Interaction (HCI) \cite{engelbart2021augmenting}, by storing them as digital files like photos or videos, may fundamentally alter how we naturally remember things—especially when AI is involved in modifying these externalized memories. This study highlights the critical responsibility of HCI researchers and practitioners in guiding the design and ethical implementation of AI technologies that could profoundly impact human cognitive functions. As AI increasingly influences how users interact with and interpret digital content, it is essential to understand how these systems can unintentionally distort memory and alter perception and recollection of reality. To summarize, this paper contributes to the discussion of false memories in the context of HCI and AI along the following dimensions:

\begin{itemize}
\item \textbf{Empirical Evidence on AI-implanted False Memories:} We report on one of the first evidence-based studies showing that AI-generated images and videos significantly increase false memory formation, with AI-generated videos of edited images having the greatest effect.
\item \textbf{Impact of AI on Memory Recollection:} We demonstrate that exposure to AI-edited media not only increases false memories but also boosts participants' confidence in these inaccurate recollections.
\item \textbf{Effect of Different AI Edit Types:} We explore how various AI edits (e.g., changes to people, objects, and environments) influence the severity of false memories, finding that edits involving people have the most pronounced effect.
\item \textbf{Applications and Implications of AI-Generated False Memories:} We discuss potential positive applications in HCI, such as therapeutic memory reframing and enhancing self-esteem, as well as negative consequences in legal (e.g., distorted eyewitness testimony), political (e.g., manipulation of public opinion), and societal (e.g., spread of misinformation) domains.
\item \textbf{Ethical Implications \& Potential Mitigation Strategies:} We examine the ethical concerns and broader implications of AI-edited media in misinformation and propose strategies, in the context of HCI, to reduce the risks of AI-induced false memories, including public awareness campaigns and ethical AI design interventions
\end{itemize}

\section{Related Work}
This study builds upon and extends a rich foundation of interdisciplinary research at the intersection of cognitive psychology, human-computer interaction, and AI. Our work synthesizes and advances knowledge from three primary domains: false memory research, the study of AI-generated misinformation, and HCI research.

\subsection{HCI and Human Memory}
HCI research has long explored ways to augment and support human memory through technology, a concept that can be traced back to Doug Engelbart's pioneering work on augmenting human intellect, where he envisioned the computer as an externalization of human cognitive processes such as intellect and memory \cite{engelbart2021augmenting}. Early work proposed wearable systems as "remembrance agents" providing contextually relevant information \cite{rhodes1997wearable}. More recent approaches have leveraged life-logging technologies to capture and retrieve personal experiences \cite{Chen_2010, kalnikaite2010now, Gurrin2014, harvey2016remembering}. Several studies have investigated digital memory aids for older adults and those with memory impairments, including prospective memory training \cite{Chan_2019, Chan_2020} and safety settings for couples \cite{McDonald_2021}. Researchers have also examined how to design for embodied remembering \cite{Overdevest_2023, 10.1145/2984751.2984776} and support autobiographical memory in depression \cite{Qu_2019}.

Beyond individual memory, research has explored participatory approaches to collective and cultural memory-making \cite{Kambunga_2020}. The emergence of virtual and augmented reality has opened up new possibilities for memory augmentation, with researchers proposing "superpowers" like enhanced recall \cite{Sadeghian_2021} and exploring AR-based memory cues \cite{Bahnsen_2024}.

Wearable devices have been developed to enhance auditory attention and memory \cite{Urakami_2023}, as well as to leverage olfactory cues for memory reactivation \cite{Amores_Fernandez_2023}. Personal knowledge management applications have been designed to support long-term memory augmentation \cite{Schneegass_2021}, while other work has investigated sound-based mementos for people with visual impairments \cite{Yoo_2024}. Recent advances in AI and large language models have enabled more sophisticated memory assistants, such as conversational agents that can infer memory needs in real-time \cite{10.1145/3613904.3642450} and AI-assisted journaling tools \cite{Kim_2024}.

Researchers have also explored gamification of memory experiments \cite{ElAgroudy_2021} and modeled language learning and forgetting \cite{Ma_2023}. As technologies for memory augmentation become more prevalent, HCI researchers continue to grapple with design challenges around user experience, privacy, and the ethical implications of outsourcing human memory to digital systems.

\subsection{AI-generated Media and Misinformation}
In recent years, the rapid advancement of AI technologies, particularly large language models \cite{chang2024survey} and generative visual models \cite{yang2023diffusion}, has led to their widespread integration into work processes and daily life. This integration raises critical questions about the potential impact of AI on human cognition, particularly in the area of misinformation (the spread of falsehoods regardless of intent) and disinformation (deliberately misleading content or propaganda).

Researchers have identified an increase in AI-generated disinformation campaigns \cite{kertysova2018artificial, goldstein2023generative} and the factors that make them disruptive to people's ability to discern true and false information \cite{goldstein2024persuasive, groh2022deepfake, groh2022human, sirlin2021digital, nightingale2022ai, lakkaraju2020fool, brown2020language, epstein2023art}. These factors include authoritative tone \cite{karinshak2023working}, persuasive language \cite{voelkelartificial, karinshak2023working, goldstein2024persuasive}, and targeted personalization \cite{tappin2023quantifying}. AI-generated content has also been shown to influence people's attitudes \cite{jakesch2022interacting, voelkelartificial, kidd2023ai}. 

The concern about AI-generated misinformation is amplified by the known yet unresolved tendency of AI models to hallucinate or generate false information, either intentionally or unintentionally \cite{huang2023survey, danry2022deceptive, zhou2023synthetic, xu2023combating}. Further, initial studies have provided evidence for the potential of AI systems to influence memory formation. In a separate study, a social robot that provided users with incorrect information before a memory recognition test had an influence comparable to that of humans. The study found that even though the inaccurate information was emotionally neutral and not inherently memorable, 77\% of the falsely provided words were incorporated into the participants' memories as errors \cite{huang2023unavoidable}.

The spread of misinformation on social media has become a major concern, prompting extensive HCI research. Studies have explored user interactions with misinformation \cite{10.1145/3313831.3376612, 10.1145/3637405}, the role of visual content \cite{10.1145/3579542, 10.1145/3613904.3642448}, and various interventions to combat it \cite{10.1145/3555637, 10.1145/3491102.3517717}.
AI's role in misinformation detection and mitigation has gained attention. Research has examined AI-based credibility indicators' effects on news perception \cite{10.1145/3555562} and the potential of using layperson judgments to combat misinformation \cite{10.1145/3313831.3376232}. 

Studies have revealed risks associated with AI-generated content \cite{10.1145/3613904.3642596} and explored AI-powered tools for fact-checking and misinformation detection \cite{10.1145/3563359.3597396}. Research has also examined the characteristics of AI-generated misinformation compared to human-created content \cite{10.1145/3544548.3581318}.

Researchers are exploring novel approaches to leverage AI in combating misinformation, including the development of intelligent tools that encourage metacognitive skills "in the wild" \cite{10.1145/3491101.3519661}.
The intersection of AI and misinformation presents both opportunities and challenges for the HCI community. Future research should focus on developing robust, transparent, and user-centered AI systems to support users in navigating the complex information landscape while addressing the implications and potential unintended consequences of AI-driven interventions.

\subsection{False Memories Research}
Research by Loftus and colleagues has established false memories as a crucial area of psychological research \cite{loftus1974reconstruction, loftus1997creating, loftus1995formation, loftus1996eyewitness, loftus1981eyewitness}. Their investigations into memory malleability and the misinformation effect have significantly influenced our understanding of memory processes, with far-reaching implications across psychology, law, and education \cite{loftus1996eyewitness, loftus1981eyewitness, loftus1975eyewitness}.

A seminal study revealed the profound impact of linguistic framing on eyewitness memory, demonstrating that the choice of verbs in questioning could markedly influence participants' speed estimates of a car accident they had witnessed \cite{loftus1974reconstruction}. The "Lost in the Mall" experiment demonstrated the feasibility of implanting entirely fabricated childhood memories \cite{loftus1995formation}. A recent replication, utilizing a larger sample size, corroborated and extended these findings, reporting a 35\% false memory rate compared to the original study's 25\% \cite{murphy2023lost}. These results not only reinforce the robustness of the initial findings but also underscore the potential ramifications for eyewitness testimony in legal contexts.

In the context of visually induced false memories, visual stimuli can generate false memories of fictitious events \cite{stephan2017visual, robin2022effects, wang2018nature, garry2005photographs}. Methods include presenting scenes with omitted elements \cite{miller1998creating}, personal photos \cite{lindsay2004true}, and narrative instructions \cite{wade2002picture}. One study found 50\% of participants developed false memories after viewing fake childhood photos and guided imagery, highlighting implications for clinical and legal professionals \cite{wade2002picture}.


Research on false memories has expanded to encompass various technological domains. The advent of immersive technologies has introduced new challenges in memory research, with studies demonstrating the occurrence of source confusion between reality and VR experiences \cite{Bonnail_2024}. Researchers have developed frameworks categorizing XR Memory Manipulations (XRMMs) based on their impact on memory processes, emphasizing the ethical concerns and potential opportunities associated with manipulating perception and memory in XR environments \cite{Bonnail_2023}. Additionally, vulnerabilities in chatbot memory mechanisms that allow for the injection of misinformation alongside personal knowledge have been demonstrated \cite{Atkins_2023}. These collective findings have fundamentally reshaped our understanding of memory processes, emphasizing the dynamic and reconstructive nature of memory.


\section{Methodology}

This study examines how AI-edited images and videos influence the formation of false memories. Specifically, we aim to assess whether exposure to AI-modified visuals affects individuals' recollections of the original scenario. In a pre-registered between-group experiment (AsPredicted \#188511 - not yet public for anonymity), 200 participants were initially shown 24 baseline images. After a filler task, they viewed AI-altered versions of the images, with changes such as adding or removing objects, altering people, or modifying environmental features, depending on their randomly assigned condition (50 participants per group). Examples include changing ethnicity, time of day, or military presence, as seen in Figure~\ref{fig:example-stimulus}. In two conditions, the static images were converted into 5-second videos using a generative AI tool. Participants then answered 24 questions, one per image, to assess their memory of the originals. The following section details the study design, experimental conditions, and protocols.

\begin{figure*}
    \centering
    \includegraphics[width=1\linewidth]{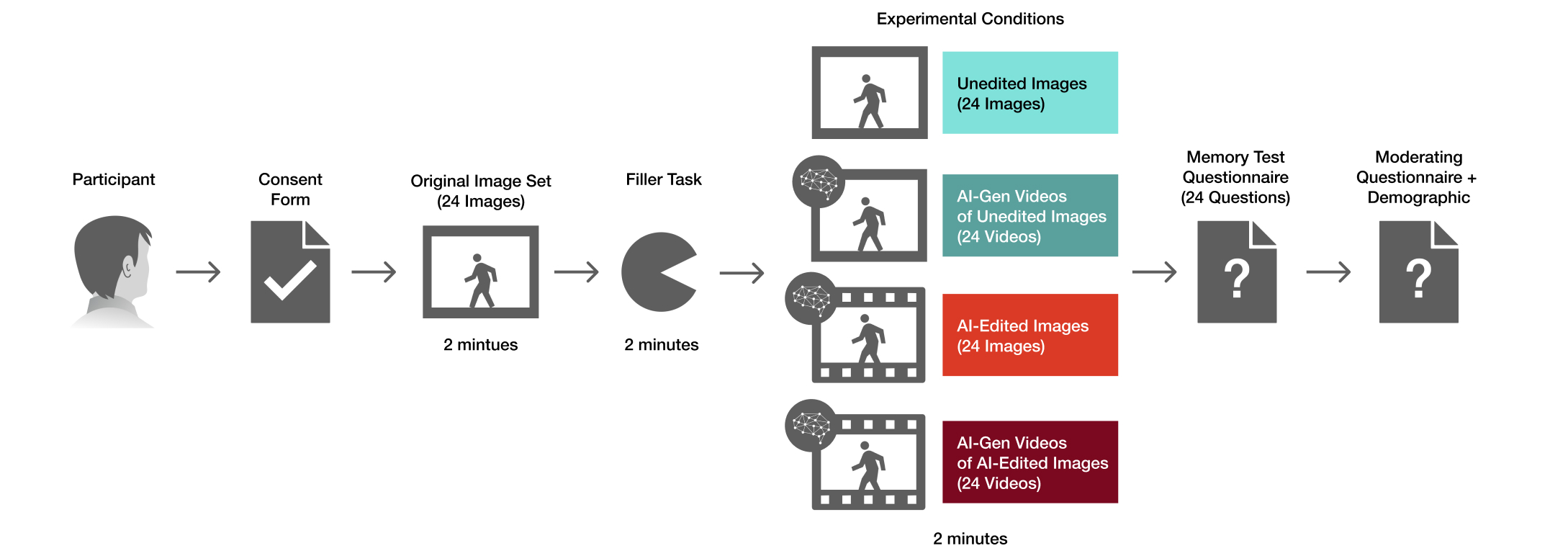}
    \caption{This figure illustrates the study procedure for our experiment examining how AI-generated images and videos can induce false memories. Participants first viewed original images to establish baseline memories, then were exposed to AI-modified versions after a filler task. These modifications included changes like increased military presence or removed climate change indicators. Finally, participants' memories of the original images were assessed through a series of questions, allowing researchers to measure the impact of AI-edited visuals on recall accuracy.}
    \label{fig:experimental-protocol}
\end{figure*}

\subsection{Study Design Rationale}
We particularly designed the study to simulate situations where individuals encounter AI-edited media in everyday contexts, particularly through social media and news platforms. In these scenarios, images are often anonymously altered or automatically edited by AI filters, frequently without the user's knowledge. The study also examines cases where others in the user's social circle, such as friends or family members, might edit personal images on behalf of the user without fully informing them of the details. In the "Future Research" section, we recommend exploring the impact of AI on false memories when people edit images themselves. We hypothesize that individuals may eventually forget they made these edits, potentially resulting in an effect similar to the scenarios described earlier. Our study’s design incorporates key elements reflective of how people currently interact with digital content, offering insights into the potential risks associated with AI-manipulated media.

\subsubsection{AI-Enhanced Media is Ubiquitous} With the increasing integration of AI-powered image and video editing tools in widely-used applications like Instagram, TikTok, and news websites, and increasingly in phone cameras themselves. For example, features like Google’s “Best Take”~\cite{besttake} can remove unwanted elements or combine the best parts of multiple shots seamlessly. Apps such as Apple Photos, Google Photos, and Samsung’s Galaxy AI offer built-in AI tools that allow users to easily edit, remove, or alter photo features with just a few simple steps. Users are frequently exposed to content that has been altered, often without their knowledge. Our study reflects this reality by introducing both static (AI-edited images) and dynamic (AI-generated videos) content, highlighting the various forms in which individuals might encounter altered depictions of events in their daily lives. Additionally, we explore scenarios where AI-generated videos are created based on AI-edited images, representing a multi-layered approach to content generation.


\subsubsection{Use of AI Labels in Media} As social media platforms begin to introduce labels indicating that content has been altered or generated by AI, our study incorporates a similar label to investigate its effectiveness. Participants in all conditions were shown images with a label indicating “enhanced image”. This feature provides an ecologically valid environment to explore whether transparency about manipulation impacts participants’ memory accuracy, offering real-world insights.

\subsubsection{Minimal Verification by Users} While in the real world, individuals have the ability to fact-check or search for original images to verify what they see, research indicates that most people do not take the time to do so~\cite{metzger2007making}. Instead, they often trust the media they encounter, especially when it appears realistic and authoritative ~\cite{metzger2013credibility, Fogg2003}. Our study mirrors this behavior by presenting participants with edited media without requiring them to verify their authenticity. This allows us to explore the extent to which unverified AI-enhanced content can distort memories, as well as the risks that come from widespread acceptance of such content without scrutiny.

\subsubsection{Confidence in False Memories}
One critical real-world issue is the cumulative effect of false memories. Once individuals have formed a false memory, such as believing a political figure said certain words, they may be more persuasive in influencing themselves and others that the memory is authentic~\cite{loftus1995formation, Loftus2005} and become more vulnerable to additional misinformation~\cite{Thorson2015, Kan2021}. Our study investigates not only whether false memories form from AI-enhanced media but also how confident individuals are in these false memories. This reflects a key concern in today’s information ecosystem, where manipulated content can distort reality on a large scale.

\subsubsection{Risk of False Memories in Sensitive Scenarios} False memories formed through AI-altered content can have dangerous consequences. For instance, an AI-edited video could falsely depict a President committing a crime, leading individuals to believe they witnessed such an event. This kind of misinformation could fuel conspiracy theories or political unrest. While our study uses examples that are representative of what individuals might realistically encounter in daily life, we deliberately avoided using misleading content that could cause harm or incite dangerous reactions. We achieved this by carefully selecting our content: avoiding distortions that could cause reputational damage, using images of political figures from countries with minimal impact on US politics, and excluding any harmful ideologies or extremist content. This approach allowed us to examine effects of AI-altered media while minimizing potential negative influences on participants or society at large.

\subsection{Stimulus Sets and Experiment Conditions}
\begin{figure*}
    \centering
    \includegraphics[width=1\linewidth]{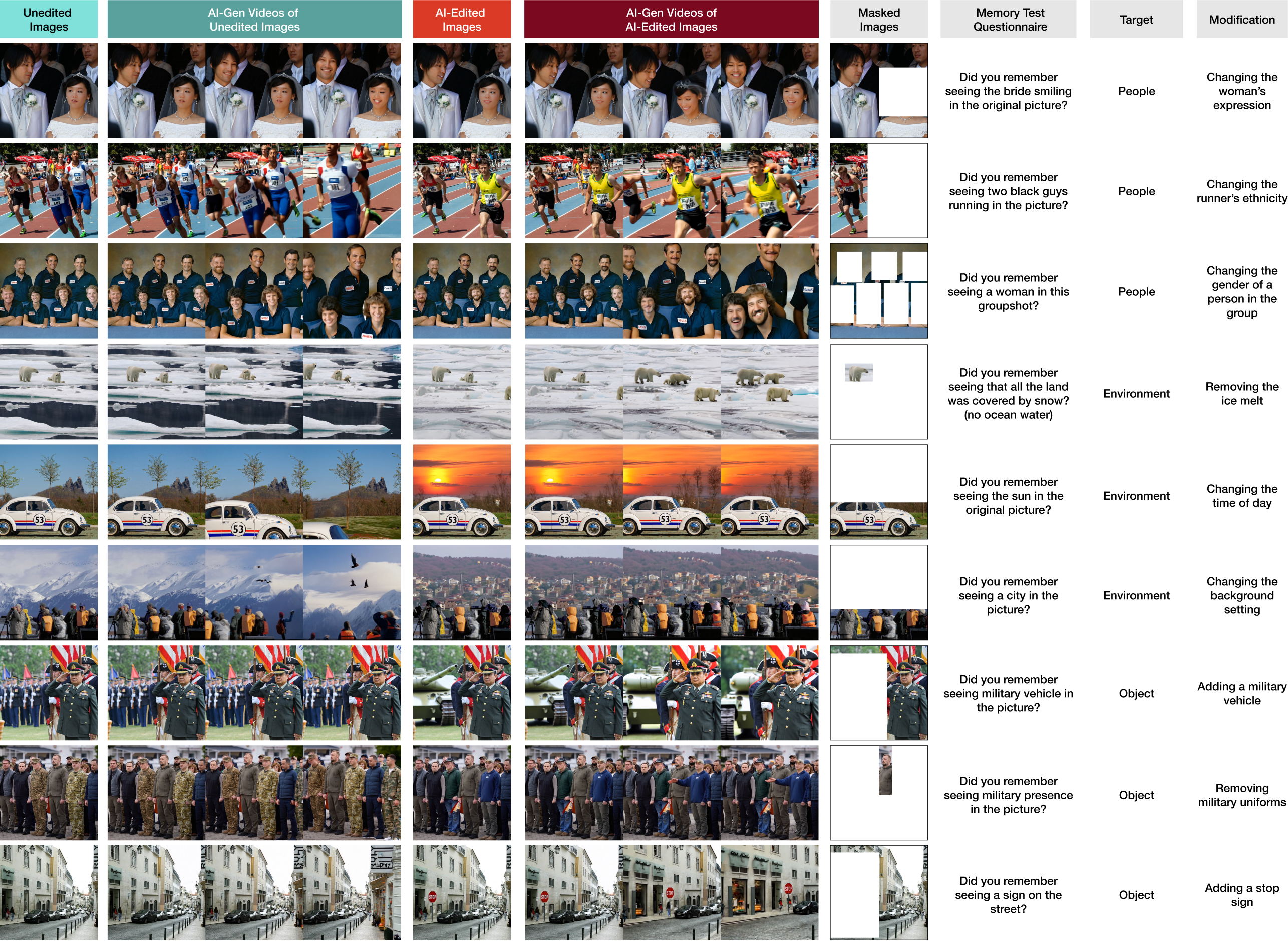}
    \caption{The stimulus set consisted of four distinct categories: unedited images, AI-edited images, AI-generated videos from unedited images, and AI-generated videos from AI-edited images. The edits were further divided into three subgroups based on the type of change: People, Objects, and Environment. In the questionnaire, masked versions of the images were used to facilitate recall without revealing the edited features.}
    \label{fig:example-stimulus}
\end{figure*}

Our study employed a diverse set of visual stimuli to investigate participants' perceptions and reactions to various types of media. The stimulus set comprised four distinct categories: unedited images, AI-edited images, AI-generated videos from unedited images, and AI-generated videos from AI-edited images:

\begin{itemize}
    \item \textbf{Control (unedited images):} Contains 24 copyright-free images curated by researchers. The image set includes political figures, such as politicians shaking hands with other leaders; personal pictures, such as views taken from a trip; and documentary pictures, such as a NASA astronaut portrait. An example of the stimulus set is shown in Figure~\ref{fig:example-stimulus}.
    \item \textbf{AI-edited images:} From the unedited set, we used Adobe Photoshop AI to edit 12 images by either removing, adding, or altering details of the pictures (The remaining 12 were left unedited as a group homogeneity check). As shown in Figure~\ref{fig:example-stimulus}, edits are categorized into 3 groups: Targeting people (changing the woman's expression, changing the runner's ethnicity, changing the gender of a person in the group), targeting environment (removing the ice melt, changing the time of day, changing the background setting), and targeting objects (adding a military vehicle, removing military uniforms, and adding a stop sign). 
    \item \textbf{AI-generated videos of unedited images:} Luma’s Dream Machine is used to convert the unedited image set into videos, primarily relying on the default auto-generation mode to create realistic motion based on the existing content in the images. For most images, no additional prompts were provided. However, for a few images where the system struggled, we added supplementary prompts to guide the model while maintaining the same intended outcome. The video length is 5 seconds. Example frames of the videos are shown in Figure~\ref{fig:example-stimulus}.
    \item \textbf{AI-generated videos of AI-edited images:} Similar to AI-generated videos of unedited images, we used Luma's Dream Machine to convert the AI-edited image set into videos. The video length is 5 seconds. Example frames of the videos are shown in Figure~\ref{fig:example-stimulus}.
\end{itemize}

Participants were initially presented with the original, unedited image set, followed by one of the stimulus sets from the four conditions. The stimuli were delivered through a customized web interface with "next" and "previous" buttons, allowing participants to navigate through all 24 visuals. However, once they completed viewing the original set, they could not go back to revisit it. The web interface was embedded within Qualtrics.

We employed a two-by-two design to investigate the influence of AI-generated images and videos on the formation of false memories. Participants were randomly assigned to one of four conditions corresponding to four sets of stimuli. This two-by-two structure allowed for a clear comparison between static and dynamic AI-altered content, as well as between unedited and AI-edited visuals, enabling us to explore how different forms of AI media affect memory distortion. 


This two-by-two structure allowed for a clear comparison between static and dynamic AI-altered content, as well as between unedited and AI-edited visuals, enabling us to explore how different forms of AI media affect memory distortion. 

\subsection{Measurement}
Participants were presented with a \textbf{Memories Test Questionnaire} consisting of 24 questions regarding their memory of the original, unedited images. Each question included a picture with the key detail of interest masked, while still providing enough context for participants to understand which image was being referenced. Example questions are shown in Figure~\ref{fig:example-stimulus}.

These questions assessed whether participants recalled specific details from the images, such as objects, people, or environmental elements (e.g., “Did you remember seeing the bride smiling in the original picture?”). The participant could either answer agree, disagree, or unsure. In addition, participants rated their confidence on a 7-point scale from 1 (extremely lacking confidence) to 7 (extremely confident). 

\begin{figure*}
    \centering
    \includegraphics[width=1\linewidth]{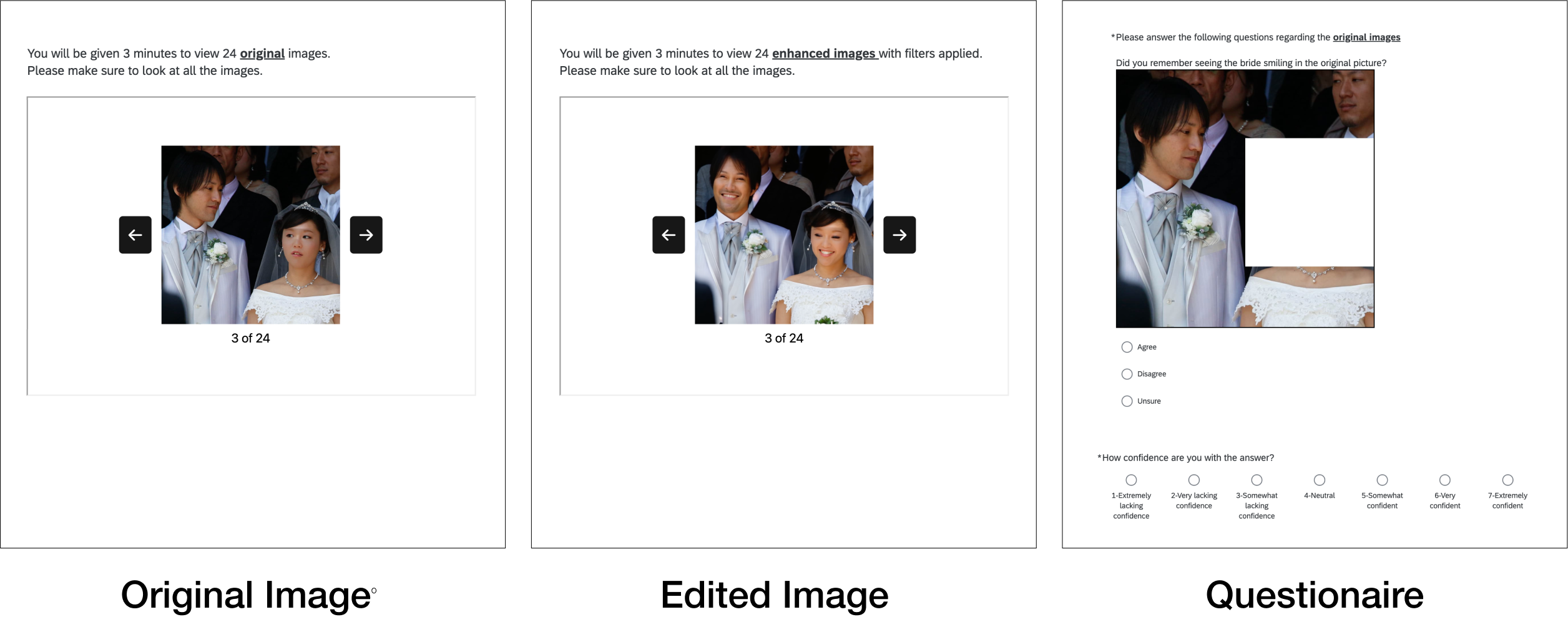}
    \caption{Survey interface components. From left to right: (1) Original image viewing instructions, (2) AI-enhanced image viewing instructions, (3) Questionnaire about original image details and confidence assessment. The sample images depict a wedding scene. The questionnaire prompts participants to recall specific details and rate their confidence in their memory.}
    \label{fig:interface}
\end{figure*}


For the moderating factors, to measure \textbf{AI Filter Familiarity}, participants indicated their familiarity with using image or AI filter technologies on a 7-point scale (1 = Not familiar at all, 7 = Very familiar). \textbf{Frequency of Forgetting}, adapted from \cite{zelinski1990memory}, assessed participants’ general memory performance, with responses ranging from 1 (Major problems) to 7 (No problems). \textbf{Memory Efficacy} was measured using a subset of items from \cite{berry1989reliability}, where participants rated their ability to remember visual and verbal information, such as recalling names and objects, on a 7-point scale (1 = Strongly disagree, 7 = Strongly agree). Finally, \textbf{Skepticism} was assessed using a scale adapted from \cite{ferrer2024naive}, where participants rated their agreement with statements reflecting distrust in official information (e.g., “The official media provides false information”) on a 7-point scale (1 = Strongly disagree, 7 = Strongly agree).

\subsection{Experiment Protocol}
The experiment was conducted using Qualtrics. A total of 200 participants were recruited from CloudResearch, with an equal 1:1 ratio of female to male participants. The study was limited to individuals residing in the United States, and participants ranged in age from 20 to 73 years old, \textit{M}=38, \textit{s.d.}=12.25. The experimental procedure began with participants signing a consent form, which included a disclosure regarding the possibility of deception in the study.

In the questionnaire's introduction, participants are informed that they will be presented with a set of images, followed by their corresponding "filtered" versions, and after observing both sets, they will be asked to provide "feedback" on the filter through a series of questions.

An initial attention check was administered to ensure focus. Participants were then shown a set of 24 unedited images for two minutes; participants could flip through the images by swiping, similar to Instagram. Following this, participants engaged in a two-minute filler task, playing Pac-Man. After the filler task, they were presented with a new set of AI-edited images labeled “AI-enhanced image.” These images featured subtle edits, though participants were not informed of the specific alterations. Participants were randomly assigned to different experimental conditions, and another attention check was administered.

To assess memory distortion, participants were then asked 24 memory test questions about their recollection of the original images. Each question was accompanied by a masked version of the image to help facilitate recall without revealing the edited features. The questions were designed to determine whether false information from the AI-edited images had been incorporated into participants’ memories. For instance, a question might ask, “Did you remember seeing military presence in the picture?” The questions were balanced between positive prompts (asking about elements that had been added) and negative prompts (asking about elements that had been removed).

Finally, participants completed a demographic questionnaire and a post-survey, which included items related to potential technical issues and moderating factors such as familiarity with AI filters, memory efficacy, tendency for forgetting, and skepticism, as well as age, gender, and education level.

\subsection{Analysis}
In order to evaluate the impact of AI-edited media on participants’ memory, we conducted 3 analyses, each designed to address different aspects of our hypothesis. The analyses were structured as follows:

\begin{enumerate}
    \item \textbf{Primary analysis} aimed to test the main hypothesis regarding the overall effect of each condition on memory. Three key metrics were observed: (1) The \textbf{Number} of reported false, uncertain, and non-false memories (i.e., how many times participants recalled incorrectly, were unsure, or recalled the original image correctly). (2) The \textbf{Confidence} levels associated with each type of memory (false, uncertain, and non-false). (3) A \textbf{Weighted score}, calculated by assigning values to each memory type (false: -1, uncertain: 0, true: 1) and multiplying them by the corresponding confidence levels, then summing these products.
    \item \textbf{Subgroup analysis} examining the number of false memory occurrences across conditions in the subgroups separated by two conditions, i.e., \textbf{image content} and \textbf{subject of edit}, to uncover any category-specific effects of generative content on memory.
    \item \textbf{Moderating Factors} exploring the effect of moderating factors on the number of false memory occurrences using a mixed-effect regression model.
\end{enumerate}

We classified participants’ memories as false, uncertain, or non-false based on their responses (agree, disagree, or unsure) to specific questions about details of the original images. If a participant answered a question incorrectly, it was marked as a false memory. A correct answer was classified as a non-false memory, and if the participant responded with “unsure,” it was categorized as uncertain.

For each test, we first assessed if the normality assumption was met for each outcome variable distribution using the Shapiro-Wilk test. If the normality assumption was not met, we performed a Kruskal-Wallis test followed by a post-hoc Dunn test using the Bonferroni error correction. If the normality assumption was met, we then conducted a homogeneity test using a Levene test to assess whether the samples were from populations with equal variances. If the samples were not homogeneous, we ran a Welch analysis of variance (ANOVA) and a Tukey's honestly significant difference test (Tukey's HSD) test. If the samples were homogeneous, we ran a basic ANOVA test with a Tukey post-hoc test.

\subsection{Approvals} 
This research was reviewed and approved by the [Anonymized] Institutional Review Board, protocol number [Anonymized]. The study was also preregistered at AsPredicted (\#188511 - not yet public for anonymity).

\section{Result of Primary Analysis}

\begin{figure*}
    \centering
    \includegraphics[width=\linewidth]{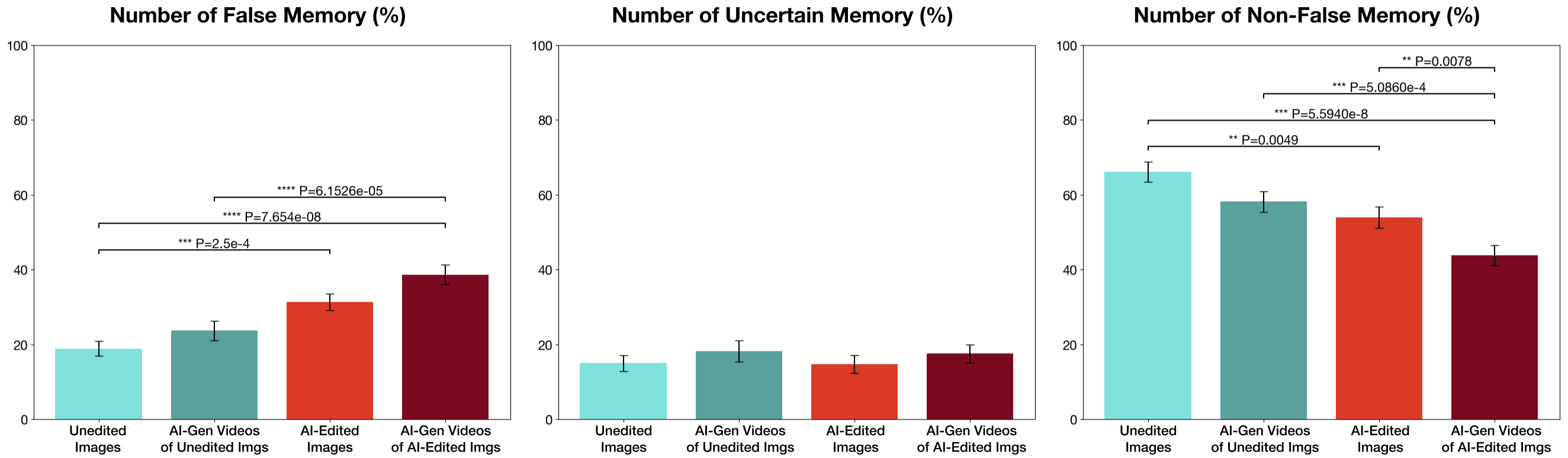}
    \caption{The percentage of reported false, uncertain, and non-false memories (i.e., how many times participants recalled incorrectly, were unsure, or recalled the original image correctly) were analyzed using a one-way Kruskal-Wallis and post hoc Dunn with FDR. P-value annotation legend: **, \textit{P}<0.01; ***, \textit{P}<0.001; ****, \textit{P}<0.0001.}
    \label{fig:number}
\end{figure*}

\noindent The results show that AI-edited images distort participants' memories of the original images, leading them to report more false memories and higher levels of confidence in the false memories, while converting the images into videos further increases number of reported false memories. The statistics are illustrated in Figure~\ref{fig:number}.

\subsection{Number of Recalled Memories by Categories}
Shapiro-Wilk tests revealed non-normality across all four experimental conditions (P-value ranging from 3.238e-5 to 5.79e-4). We employed one-way Kruskal-Wallis tests to analyze variation.

\subsubsection{Number of False Memories}
As illustrated in the first column of Figure~\ref{fig:number}, the test indicates significant differences in the number of reported false memories between the conditions, \textit{H}=34.157, \textit{P}=1.836e-07, \textit{P}<.05. The AI-edited images induced significantly more false memories than the unedited images (1.67x), while AI-generated videos of AI-edited images amplify the number of false memories leading to a 2.05x number of reported false memories. However, AI-generated videos from unedited images result in 1.25x compared to control. Statistics: control, \textit{M}=18.878, \textit{s.d.}=14.164, number of reported false memories out of 12 (\textit{\#})=2.265; AI-gen videos of unedited images: \textit{M}=23.667, \textit{s.d.}=18.055, \textit{\#}=2.840; AI-edited images: \textit{M}=31.373, \textit{s.d.}=15.965, \textit{\#}=3.765; AI-gen videos of AI-edited images: \textit{M}=38.667, \textit{s.d.}=18.838, \textit{\#}=4.640. post hoc Dunn test with Benjamini–Hochberg (FDR) correction: control vs AI-edited images, \textit{P}=2.5e-4; control vs AI-gen videos of AI-edited images, \textit{P}=7.654e-08; AI-gen videos of unedited images vs AI-gen videos of AI-edited images, \textit{P}=6.153e-05.

\subsubsection{Number of Uncertain and Non-false memories}
There were no significant differences in the number of reported uncertain memories between conditions as shown in the second column of Figure~\ref{fig:number}, \textit{H}=0.903, \textit{P}=8.246, \textit{P}>0.05. Meanwhile, in the third column, the test indicates significant differences in the number of reported non-false memories between the conditions with, \textit{H}=30.448, \textit{P}=1.111e-06, \textit{P}<.0071. All three edited sets introduce the lower number of reported non-false memories, i.e. 0.88x, 0.82x, 0.67x, in AI-gen videos of unedited images, AI-edited images, and AI-gen videos of AI-edited images conditions respectively, compared to control . Statistics: control, \textit{M}=66.156, \textit{s.d.}=18.929, number of reported non-false memories out of 12 (\textit{\#})=7.939; AI-gen videos of unedited images: \textit{M}=58.167, \textit{s.d.}=19.543, \textit{\#}=6.980; AI-edited images: \textit{M}=53.922, \textit{s.d.}=20.032, \textit{\#}=6.471; AI-gen videos of AI-edited images: \textit{M}=43.833, \textit{s.d.}=18.545, \textit{\#}=5.260. post hoc Dunn test with Benjamini–Hochberg (FDR) correction: control vs AI-edited images, \textit{P}=0.005; control vs AI-gen videos of AI-edited images, \textit{P}=5.594e-8; AI-generated videos of unedited images vs AI-gen videos of AI-edited images, \textit{P}=5.086e-4; AI-edited images vs AI-gen videos of AI-edited images, \textit{P}=0.0078.

\subsection{Confidence in Recalled Memories}
\begin{figure*}
    \centering
    \includegraphics[width=\linewidth]{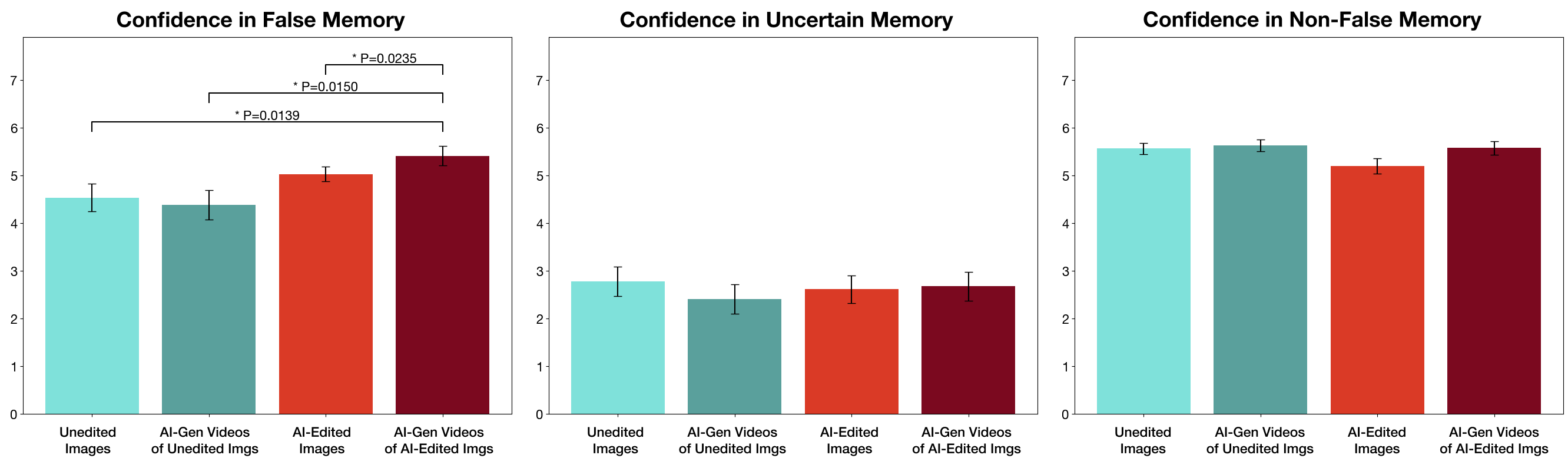}
    \caption{The confidence of recalled memories in three categories (false/uncertain/true) were analyzed using a one-way Kruskal-Wallis and post hoc Dunn with FDR. P-value annotation legend: *, \textit{P}<0.05.}
    \label{fig:confidence}
\end{figure*}

Shapiro-Wilk tests revealed non-normality across all four experimental conditions (P-value ranging from 6.709e-12 to 1.167e-10). We employed one-way Kruskal-Wallis tests to analyze variation. The results are illustrated in Figure~\ref{fig:confidence}

\subsubsection{Confidence of False Memories}
As shown in the first column of Figure~\ref{fig:confidence}, the results indicate significant differences between the conditions, \textit{H}=8.581, \textit{P}=0.0354, \textit{P}>0.05. AI-gen videos of AI-edited images and AI-edited images respectively induce 1.19x and 1.1x increase in the level of confidence in false memories vs control. Statistics: control, \textit{M}=4.536, \textit{s.d.}=2.041; AI-gen videos of unedited images: \textit{M}=4.383, \textit{s.d.}=2.220; AI-edited images, \textit{M}=5.027, \textit{s.d.}=1.092; AI-gen videos of AI-edited images: \textit{M}=5.412, \textit{s.d.}=1.449. post hoc Dunn test with Benjamini–Hochberg (FDR) correction: control vs AI-gen videos of AI-edited images, \textit{P}=0.0139; AI-gen videos of unedited images vs AI-gen videos of AI-edited images, \textit{P}=0.0150; AI-edited Images vs AI-gen videos of AI-edited images, \textit{P}=0.0235.

\subsubsection{Confidence in Uncertain and Non-False Memories}
The tests indicate no significant differences in the confidence in uncertain and non-false memories, the numbers are illustrated in the second and third column of Figure~\ref{fig:confidence}. Statistics: Uncertain Memories, \textit{H}=0.726, \textit{P}=0.867, \textit{P}>0.05; Non-false Memories, \textit{H}=5.012, \textit{P}=0.170, \textit{P}>0.05.

\subsection{Weighted Score}
We calculated the weighted score by multiplying the score assigned to each memory type (\textit{False} = -1, \textit{Uncertain} = 0, \textit{Non-False} = 1) by the participant’s confidence level for that memory (ranging from 1 to 7). The Shapiro-Wilk test results for all groups show \textit{P}<0.05, indicating that the data in each group is normally distributed. The Levene test result (\textit{P}=0.683) suggests homogeneity of variances across the groups. The one-way ANOVA test yielded a highly significant result (\textit{F}=14.577, \textit{P}=1.312e-08, \textit{P}<0.007), indicating substantial differences among the group means. Statistics: control, \textit{M}=32.061, \textit{s.d.}=21.989; AI-gen of unedited images, \textit{M}=24.760, \textit{s.d.}=22.801; AI-edited images: \textit{M}=15.803, \textit{s.d.}=21.476; AI-gen videos of AI-edited images, \textit{M}=3.040, \textit{s.d.}=24.933.

\begin{figure}[h]
    \centering
    \includegraphics[width=1\linewidth]{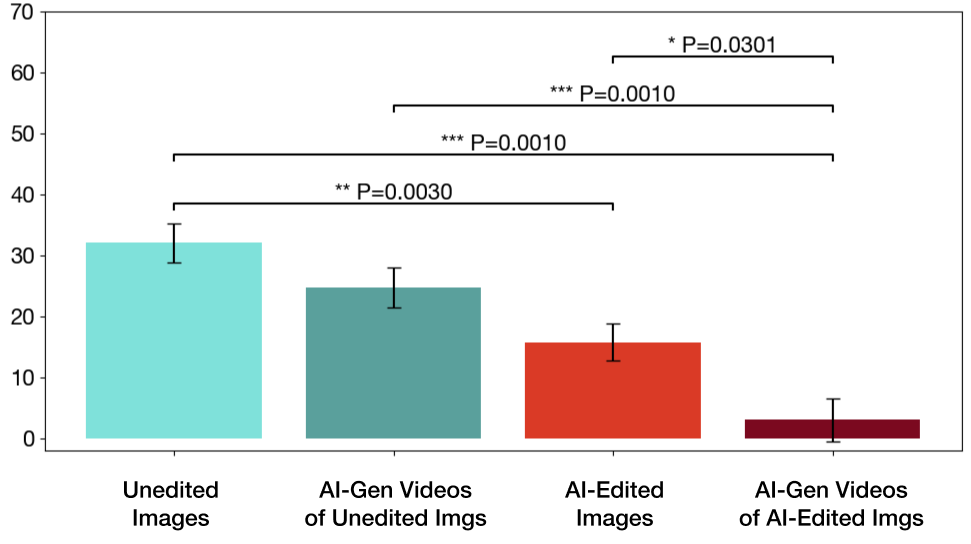}
        \caption{The weighted score of recalled memories based on number and confidence was analyzed using a one-way ANOVA and post hoc Tukey. P-value annotation legend: *, \textit{P}<0.05; **, \textit{P}<0.01; ***, \textit{P}<0.001.}
    \label{fig:score}
\end{figure}

The subsequent Tukey's post-hoc test revealed significant differences between several pairs of groups, particularly between the control and both AI-edited images (\textit{P}=0.003) and AI-gen videos of AI-edited images groups (\textit{P}=0.001), as well as between AI-gen videos of unedited images and AI-gen videos of AI-edited images groups (\textit{P}=0.001), and between AI-edited images and AI-gen videos of AI-edited images (\textit{P}=0.03). These findings suggest that the AI editing interventions, especially those involving static AI image edits and combination with AI video generation, have a significant impact on the measured variable compared to the control condition.

\subsection{Group Homogeneity Checks for Primary Analysis}
To ensure the homogeneity of our test groups, we conducted a group homogeneity check. In the test set design, half of the images remained unedited, even if they were part of AI-edited groups. For example, in the AI-edited images group, 12 out of 24 images were left unedited and matched those in the control group. The same approach was applied to the AI-generated videos of the AI-edited images group. We hypothesized that these unedited sets across groups would produce similar results in terms of the number and confidence of false, uncertain, and non-false memories across the three experimental conditions.

Following the same procedure as our main analysis, we conducted statistical tests. The Shapiro-Wilk test results for normality were as follows: control (\textit{W}=0.934, \textit{P}<0.001), AI-gen videos of unedited images (\textit{W}=0.935, \textit{P}<0.001), AI-edited images (\textit{W}=0.937, \textit{P}<0.001), and AI-gen videos of AI-edited images (\textit{W}=0.942, \textit{P}<0.001). Kruskal-Wallis tests resulted in P-values ranging from 0.425 to 0.927, indicating no significant differences between the groups. As hypothesized, we found no evidence that the unedited sets across groups produced differing results across the three experimental conditions. This finding supports the validity of our experimental design. The results of the Kruskal-Wallis test are shown in Table~\ref{tab:homogeneity}.

\begin{table}[h]
\centering
\begin{tabular}{lrr}
\hline
 Condition & H Statistic & P-Value \\
\hline
Number of False Memories & 1.359 & 0.715 \\
Number of Uncertain Memories & 1.961 & 0.581 \\
Number of Non-False Memories & 0.502 & 0.918 \\
\hline
Confidence in False Memories & 1.073 & 0.784 \\
Confidence in Uncertain Memories & 2.791 & 0.425 \\
Confidence in Non-False Memories & 0.460 & 0.928 \\
\hline
\end{tabular}
\caption{Group Homogeneity Checks for Primary Analysis Using Kruskal-Wallis Tests}
\label{tab:homogeneity}
\end{table}

\section{Results of Additional Analysis}

\subsection{Number of false memories based on different types of image contents}

\begin{figure*}[h]
    \centering
    \includegraphics[width=1\linewidth]{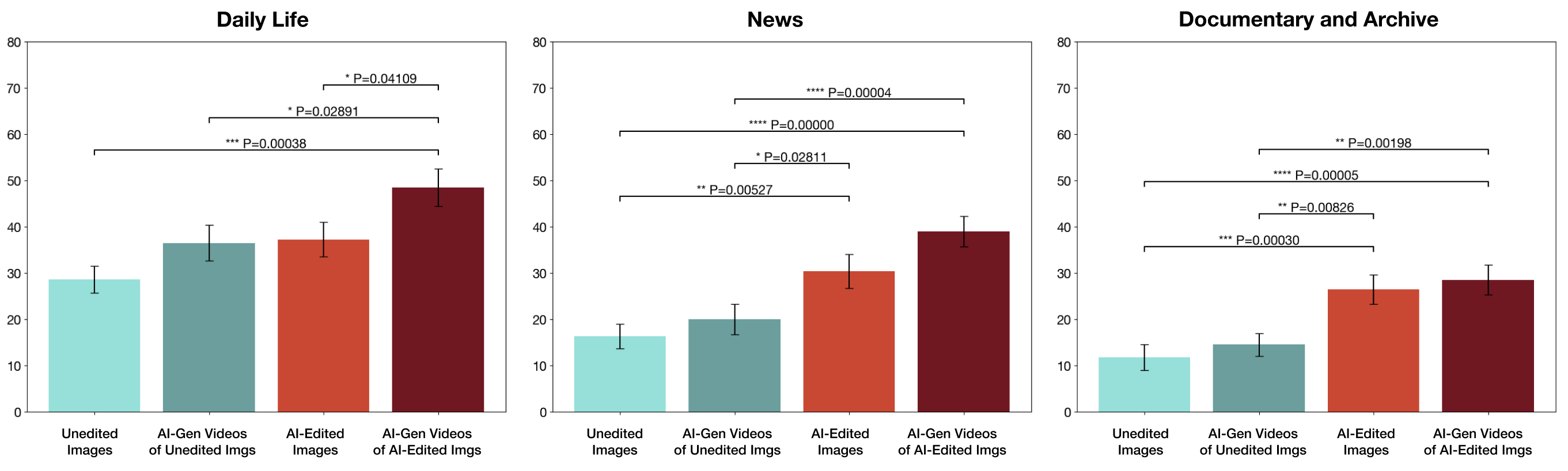}
    \caption{The number of reported false memories in three subgroups were analyzed using a one-way Kruskal-Wallis and post hoc Dunn with FDR. P-value annotation legend: *, \textit{P}<0.05; **, \textit{P}<0.01; ***, \textit{P}<0.001; ****, \textit{P}<0.0001.}
    \label{fig:group}
\end{figure*}

\noindent The stimulus set is categorized into three distinct subgroups, each representing different domain topics in visual media. Daily life photos include everyday scenes, activities, and objects that people encounter regularly, representing common visual experiences in familiar settings. News images comprise photographs typically found in news media, such as sports events and political figures. Documentary and archival materials encompass images of historical events or figures, and documentary-style shots of cultures or natural phenomena. Each of these subgroups in our test set contains 4 images, resulting in a total of 12 images across the three categories.  We apply the main analysis procedure to each of these subgroups based on number of reported false memories, aiming to evaluate the consistency of AI effects across these diverse content types. As illustrated in Figure~\ref{fig:group}, the results demonstrate a consistent pattern in the effects of AI-generated content. Statistics: Daily Life, \textit{H}=12.884, \textit{P}=0.00489, \textit{P}<0.005; News, \textit{H}=27.712,
\textit{P}=4.174e-06=\textit{P}<0.05; Documentary and Archive, \textit{H}=23.617, \textit{P}=3.003e-05, \textit{P}<0.0167.

In all three groups, there is a clear trend of increasing false memories from unedited images to AI-generated videos of AI-edited images. This similarity in trends suggests that the impact of AI-generated and AI-edited content on memory recall is robust across different types of visual stimuli, whether they depict everyday scenes, current events, or historical content, indicating that the observed effects of AI manipulation on memory are not limited to a specific type of visual content but appear to be a more general phenomenon.

\subsection{Number of false memories based on type of subject edited by AI}

\noindent As part of an exploratory analysis with descriptive statistics, we categorized the edited content into three subgroups based on the subject of the edit: people (alterations to facial features, changes in race or ethnicity, and adjustments to body shape or posture), objects (the addition or removal of items from scenes, or changes to existing objects within the image), and environments (modifications to backgrounds, lighting, weather conditions, or architectural elements). The number of false memory reports across these different categories of AI-generated edits showed a consistent pattern of increasing false memories. Yet, the impact of AI manipulation varies across the different subjects of edit. While people-related edits consistently produced the highest absolute number of false memories, environmental edits showed the most dramatic increase in false memory reported due to image edits, especially in the most complex condition of AI-generated videos of AI-edited images (2.2x of control in AI-gen videos of AI-edited images condition).

\begin{figure}[h]
    \centering
    \includegraphics[width=1\linewidth]{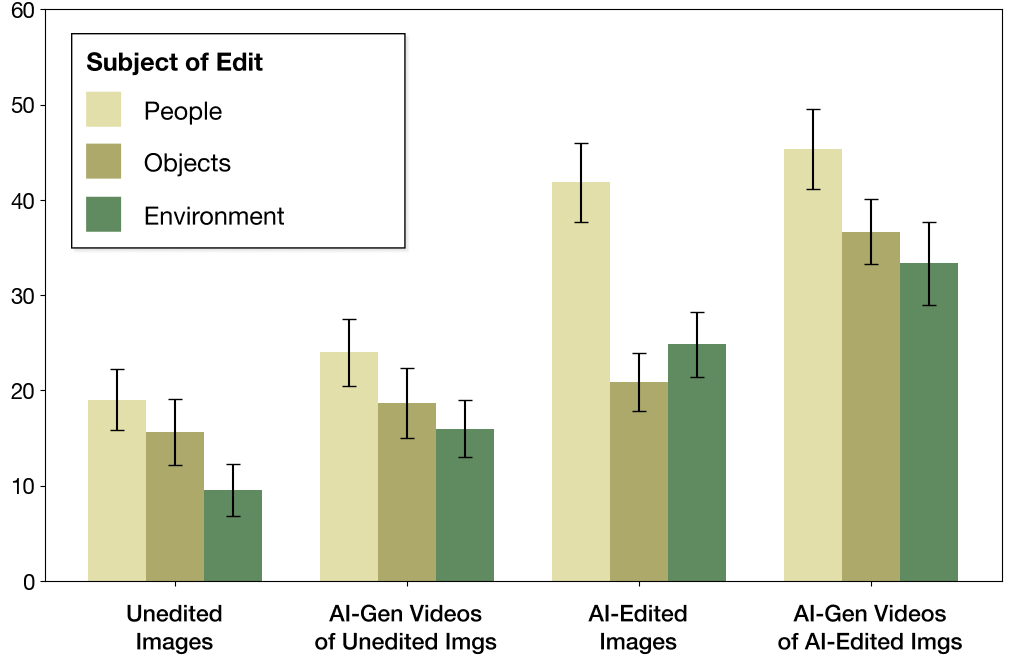}
    \caption{The percentages of reported false memories by subjects of edit.}
    \label{fig:specific}
\end{figure}


Statistics: \textbf{People}: control, \textit{M}=19.048, \textit{s.d.}=22.335; AI-gen videos of unedited images, \textit{M}=24.000, \textit{s.d.}=24.980; AI-edited images, \textit{M}=41.830, \textit{s.d.}=29.405; AI-gen videos of AI-edited images, \textit{M}=45.333, \textit{s.d.}=29.635. \textbf{Objects}: control, \textit{M}=15.646, \textit{s.d.}=24.376; AI-gen videos of unedited images, \textit{M}=18.667, \textit{s.d.}=25.957; AI-edited images, \textit{M}=20.915, \textit{s.d.}=21.854; AI-gen videos of AI-edited images, \textit{M}=36.667, \textit{s.d.}=24.267. \textbf{Environment}: control, \textit{M}=9.524, \textit{s.d.}=19.048; AI-gen videos of unedited images, \textit{M}=16.000, \textit{s.d.}=21.333; AI-edited images, \textit{M}=24.837, \textit{s.d.}=24.560; AI-gen videos of AI-edited images, \textit{M}=33.333, \textit{s.d.}=30.551.

\subsection{Moderating Factors}

We employed a mixed effects regression approach to investigate how variables such as gender, age, education, familiarity with AI-filter technology, and cognitive factors (frequency of forgetting and memory efficiency) relate to number of false memories. Table~\ref{tab:regression} provides the result of the regression model. The result suggests that age has a significant negative relationship (\textit{P}=0.01) with the number of false memories, yet the effect is small (\textit{Coef.}=-0.031). Meanwhile, other factors do not show significant relationships in the model.
\begin{table*}[htbp]
\centering
\begin{tabular}{lrrrrrr}
\hline
 & Coef. & Std.Err. & z & P>|z| & [0.025 & 0.975] \\
\hline
Familiarity with AI-Filter Technology & 0.040 & 0.087 & 0.454 & 0.650 & -0.131 & 0.211 \\
Age & -0.031 & 0.012 & -2.583 & 0.010*& -0.055 & -0.008 \\
Gender (Male compared to Female) & -0.040 & 0.291 & -0.138 & 0.891 & -0.610 & 0.530 \\
Gender (Other compared to Female) & 0.021 & 2.063 & 0.010 & 0.992 & -4.022 & 4.065 \\
Education Level & 0.103 & 0.119 & 0.862 & 0.389 & -0.131 & 0.336 \\
Cognitive Load & -0.133 & 0.119 & -1.121 & 0.262 & -0.366 & 0.100 \\
Memory Efficiency & -0.069 & 0.058 & -1.194 & 0.232 & -0.182 & 0.044 \\
Skepticism toward Media and Institutions & -0.021 & 0.015 & -1.376 & 0.169 & -0.051 & 0.009 \\
\hline
\end{tabular}
\caption{The mixed effects regression model result. P-value annotation legend: *, \textit{P}<0.05}
\label{tab:regression}
\end{table*}

\section{Discussion}

\subsection{AI-Edited Content Boosts False Memories with Alarming Confidence}
Our findings demonstrate the impact of AI-edited and AI-generated media on human memory distortion. Participants exposed to AI-altered images exhibited a markedly higher propensity to report false memories compared to those who viewed unedited control images. This effect was even more pronounced when participants were presented with AI-generated videos based on AI-edited images, suggesting that dynamic AI-edited  media significantly amplify the distortion of memory, effectively embedding false details deeper into participants' recollections.

Perhaps the most disconcerting aspect of the study's results is the high degree of confidence participants reported in their inaccurate recollections. The AI-edited images not only led to the formation of false memories but also instilled a misplaced sense of certainty in these fabricated recollections. This effect is maximized with AI-generated videos of edited images, which caused the most significant increase in both false memory formation and the associated confidence levels.

The implications of these findings are far-reaching and potentially alarming. The combination of false memories and high confidence levels creates a particularly dangerous scenario, as individuals are more likely to believe and act upon incorrect information they perceive as true. This phenomenon could have consequences in various contexts, from eyewitness testimonies in legal proceedings to the spread of misinformation in social and political spheres. Moreover, the study raises important questions about the nature of memory itself and how easily it can be manipulated by advanced AI technology. In the broader sense, our study shows that "externalizing" our memories by storing them digitally can change how we naturally remember things, especially when AI is involved in enhancing or altering these externalized memories.  

\subsection{The Impact of Different Types of Edits: People, Environment, and Objects}
This study explored the impact of different types of AI-generated edits—specifically changes to people, environments, and objects, on the formation of false memories. The pattern in the results suggests that while people-related details are generally harder to recall, possibly because of their complexity, they are less susceptible to additional distortion from AI manipulation. Environmental manipulations, such as altering the time of day or modifying landscape elements, proved particularly potent in distorting recall when subjected to AI editing and video generation. Object-related edits fell between these two extremes. The data underscores the varying susceptibility of different image elements to AI-induced false memories, with contextual/environmental changes showing the most striking relative increase. These findings highlight the complex interplay between AI manipulation techniques and human memory across different aspects of visual information, suggesting that while human subjects may be naturally memorable (and thus more prone to false memories), environmental details are more vulnerable to AI-induced distortions.

\subsection{Moderating Factors}
Interestingly, the mixed-effects regression model shows that age has a statistically significant negative effect on AI-induced false memories, indicating that younger individuals are more susceptible to such influences. This finding warrants further investigation into the underlying mechanisms. One possible explanation could relate to the higher exposure of younger individuals to technology, including AI-filtered media. The mere exposure effect suggests that people tend to develop a preference for things they are more familiar with~\cite{chan2021kinvoices}, which could lead to increased trust or decreased skepticism toward AI-generated content among younger users. However, with older adults increasingly adopting technology and becoming more proficient in its use~\cite{kakullaolder, faverio_oldertechusers}, we may also observe a parallel reduction in skepticism among older users in the near future. Another explanation could involve the different modes of interaction with technology across age groups. Younger individuals may allocate their attention differently when consuming information, potentially making them more susceptible to false memory formation. Nonetheless, it is crucial to emphasize that while the age-related difference is statistically significant, the effect size is relatively small, which suggests caution against overstating its practical implications.

Familiarity with AI-filter technology, while positively correlated, does not have a statistically significant moderating effect, suggesting that mere exposure or understanding of the technology may not provide protection against false memory implantation. This aligns with the earlier explanation of the inverse relationship between familiarity and skepticism, wherein increased trust may lead to greater susceptibility to false memories. Other variables, such as education level, cognitive load, and memory efficiency, do not show significant moderating effects, meaning these individual differences do not strongly influence the likelihood of developing false memories in response to AI-edited media. Interestingly, skepticism toward media and institutions, though negatively correlated, also lacks statistical significance, suggesting that even individuals with a more critical view of media may still be vulnerable to AI-induced memory distortions. This suggests that false memory formation operates beyond a simple “trust vs. mistrust” framework~\cite{orenstein2022eriksons, Sneed2006}: anyone, regardless of their background or expertise, could potentially fall victim to false memories when exposed to altered media. In practical terms, this means that AI-altered media presents a challenge to all segments of society. Educational campaigns may need to focus not just on technological literacy but also on promoting awareness of how our cognitive systems process and reconstruct information, making us all prone to forming false memories in certain contexts.

\subsection{The Impact of Label}
The study presented all AI-edited visuals with a label indicating that the content had been altered by AI, similar to the notification systems that some social media platforms have begun to implement to inform users about AI-generated or edited content. Despite these labels, we observed a significant increase in the formation of false memories in participants exposed to the AI-edited conditions. This finding aligns with prior research which has demonstrated that passive notifications or labels alone may be insufficient to mitigate cognitive biases induced by manipulated media.

For instance, previous studies \cite{10.1145/3313831.3376232}, have shown that prompting users to actively consider the accuracy of social media posts can reduce the likelihood of sharing misinformation online. This suggests that labels need to go beyond mere informational disclosure. Instead of serving as passive indicators, labels should be designed to actively engage users in a more thoughtful, reflective process regarding the content they are viewing.

One potential explanation for the persistence of false memories despite labeling is that such notifications, while informative, do not sufficiently alter the cognitive processing of the visual material. Human memory is inherently reconstructive, and when individuals are presented with altered content—especially content that seems plausible or familiar—they may integrate it into their existing memory frameworks, regardless of the presence of a label. Simply knowing that content has been edited may not prevent the automatic processing of memory integration and reconstruction, which can lead to false memories.

To effectively reduce the impact of AI-edited media on memory, a more proactive approach to labeling may be necessary. This could involve enhancing the salience of the labels, providing users with interactive or cognitive prompts to reflect on the accuracy or authenticity of the content, or even incorporating reminders throughout the interaction with AI-edited material. Such interventions would shift the role of labeling from a passive alert system to an active cognitive tool that helps users critically engage with the content.

\subsection{Implications of AI-implanted False Memories and Human-Computer Interaction}
Our study has shown that AI has the capacity to manipulate human memory by altering images or videos, creating false recollections that may not align with reality. In Human-Computer Interaction, the implications of these AI-implanted false memories are significant, posing both challenges and opportunities, as shown in figure~\ref{fig:false-memory-example}. In this section, we explore both the negative and positive implications of AI-implanted false memories in HCI, as well as potential mitigation strategies for each.

\begin{figure*}
    \centering
    \includegraphics[width=1\linewidth]{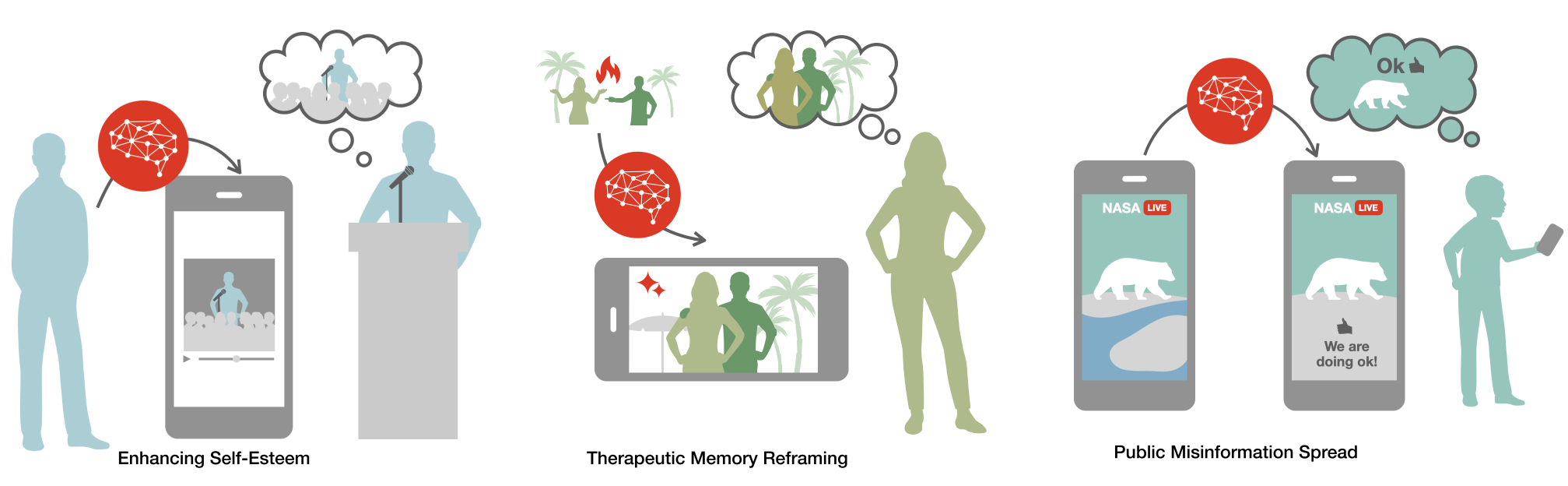}
    \caption{Examples of the potential implications of AI-generated media. The first scenario, “Enhancing Self-Esteem,” shows how AI can be used to improve personal memories, boosting confidence by enhancing positive recollections. The second scenario, “Therapeutic Memory Reframing,” illustrates the use of AI to alter memories of past events, reframing distressing moments in a more positive light for therapeutic purposes. The third scenario, “Public Misinformation Spread,” demonstrates how AI can manipulate information in public media. For instance, a student may have seen an accurate NASA climate change video in a classroom, but later, their memory could be updated with an AI-generated false video from social media showing no evidence of climate change, altering their perception of the issue.}
    \label{fig:false-memory-implication}
\end{figure*}

\subsubsection{Wrongful Legal Accusations}
In the legal domain, AI-generated false memories pose a significant and potentially life-altering risk. Altered media could falsely implicate individuals in crimes they did not commit, leading to a cascade of negative consequences. When AI is used to generate or manipulate this type of media, the potential for distorting public perception and influencing key figures in the legal process—such as witnesses—becomes even greater.

Elizabeth Loftus, a psychologist who pioneered false memory research, demonstrated the potential consequences of memory distortion in real-world cases. She highlighted the case of Steve Titus, who was wrongfully convicted of rape in 1980 \cite{wixted2021test}. Titus was identified by the victim with low confidence during a photo lineup, but by the time of the trial, her confidence had increased due to suggestive identification procedures and additional police information. Despite this shaky identification, Titus was convicted. Months later, new evidence pointed to the actual culprit, and the witness realized her mistake. Although Titus was eventually exonerated, the wrongful conviction caused financial and emotional ruin, leading to his untimely death at age 35, just before his lawsuit against the police was to be heard.

Today, news headlines are rapidly circulated and amplified through memes, posts, and viral content, which often simplify or sensationalize events. When AI is used to generate or manipulate this type of media, the potential for distorting public perception and influencing key figures in the legal process—such as witnesses—becomes even more extreme. An AI-generated meme that exaggerates or fabricates details of a crime scene could rapidly spread on social media platforms, creating a distorted narrative. Even a subtle alteration to an image or video, like placing someone in a misleading context or falsely associating them with a crime, can easily go viral. This altered media could impact not just the general public’s view of the case but also affect the memory of potential witnesses and jury members. Witnesses who are exposed to AI-generated memes or fake posts might unwittingly internalize the fabricated details, incorporating them into their own recollections of the event. What starts as a vague memory could solidify into false certainty, influenced by the repeated exposure to such misleading content.

This is especially dangerous in a courtroom setting where witness testimony often plays a pivotal role in the outcome of cases. If a witness’s memory has been unknowingly shaped by AI-generated online content, their testimony may seem credible but could be completely inaccurate. As our study has shown, in cases where the witness originally had little confidence or memory of the event, exposure to false narratives via social media could inflate their confidence and alter their recall, much like what happened in the Steve Titus case but amplified by the reach and power of AI-edited media. Given that the use of digital media is increasingly common in legal settings, systems that handle evidence, such as courtroom presentation tools or police body camera footage, must integrate robust AI-detection mechanisms. These mechanisms should be capable of preventing AI-altered media from being used as evidence in court, ensuring that AI-augmented interactions with digital content do not contribute to miscarriages of justice.

\subsubsection{Public Misinformation Spread}
One of the most pernicious and potentially damaging consequences of AI-edited false memories is the ability to manipulate public perception of historical or societal events on a grand scale. For instance, AI could be employed to alter an image or video footage of a peaceful protest by introducing elements such as military presence or violent altercations. This manipulation could cause individuals who view the altered media to falsely remember the event as violent or aggressive, even if they previously saw the original video or were present at the event. 


For example, during the 2024 presidential election campaign, significant attention is given to crowd sizes at political events~\cite{harvardRealNumbers, politico}, along with concerns about AI being used to manipulate images~\cite{coloradoImagesAbound, nytimesDespiteTrumps}. In the context of false memories, a person who attended one of these events may later encounter AI-edited images or videos on social media that subtly alter the crowd size or atmosphere. After seeing these modified visuals, their original memory of the event may shift, leading to a distorted recollection of what they truly experienced. This altered perception could cause them to believe the candidate is more popular or influential than they actually are, or that there was greater support for a particular policy than there was in reality. As a result, individuals might align their opinions with what they perceive as the majority view, potentially influencing voter preferences and election outcomes based on manipulated or exaggerated depictions.

The ramifications of such misinformation are profound and far-reaching. It could lead to widespread public confusion, distorting not only individual memories but also our collective understanding of historical events. This manipulation of shared memories has the potential to polarize societal beliefs and exacerbate existing social divisions, especially when disseminated widely on social media platforms and other digital channels. Moreover, the proliferation of AI-edited media is eroding trust in visual content, leading to increased skepticism towards all images and videos, including authentic ones.

In the context of HCI, this type of misinformation underscores the critical need for sophisticated interfaces that can help users differentiate between authentic media and AI-generated or AI-altered content. For instance, a study has shown that a notification system that primes people to focus on the accuracy of social media posts could help reduce the sharing of misinformation online \cite{pennycook2021shifting}.

\subsubsection{Reinforcement of Biases and Stereotypes}
AI-generated false memories can reinforce existing biases and stereotypes by subtly altering images or videos to fit preconceived notions. For example, AI editing could inadvertently modify media to emphasize negative stereotypes about specific groups, which can lead to false memories that perpetuate harmful narratives. This is particularly dangerous in the context of racial, gender, or cultural bias, where AI-manipulated media could fuel prejudice and discrimination.

The reinforcement of biases through AI-generated false memories is a critical concern that intersects with issues of social justice and equality. AI systems, if not carefully designed and monitored, can inadvertently perpetuate and amplify societal biases. For instance, an AI system trained on biased data might systematically alter images to conform to stereotypical representations of certain groups, leading to a feedback loop that reinforces these biases in human memory and perception.
This issue presents a significant challenge for HCI designers and researchers. How can we create systems that not only avoid reinforcing biases but actively work to counteract them? One approach could be to develop AI systems that are specifically trained to detect and flag content that may reinforce harmful stereotypes. These systems could be integrated into content moderation tools for social media platforms and other digital spaces where user-generated content is shared. 

Additionally, HCI designers should consider implementing features that promote diversity and inclusion in digital spaces \cite{bursztein2024leveraging, pillis2024ai, vrdiversity}. This could include algorithms that ensure a diverse representation in image feeds, or tools that help users critically examine their own potential biases when interacting with digital content.

\subsubsection{Mitigating AI-implanted False Memories}
Addressing the challenges posed by AI-implanted false memories requires a multifaceted approach that combines technological solutions, policy measures, and public education initiatives. To combat the spread of public misinformation, AI-detection tools should be developed and integrated into social media platforms and news aggregators. These tools would actively scan and flag potentially manipulated content, alerting users to the possibility of false information. Additionally, media literacy programs should be expanded to educate the public on the prevalence of AI-altered media and the importance of critical evaluation of online content.

In the political sphere, stricter regulations should be implemented regarding the use of AI-generated or AI-altered content in political campaigns. Political organizations and candidates should be required to disclose any use of AI in their campaign materials. 

Furthermore, professionals in the legal field, including judges, lawyers, and law enforcement officers, should receive comprehensive training on recognizing potential signs of AI manipulation in digital evidence. This training should be regularly updated to keep pace with advancements in AI technology. By adopting these mitigation strategies, the risks posed by AI-generated false memories in critical areas, such as public misinformation, political manipulation, and legal justice, might be minimized. However, more research is needed to fully understand the extent of AI’s influence on human memory and behavior, especially in high-stakes situations. There is a pressing need for interdisciplinary studies that bring together experts in psychology, HCI, and AI to explore how automated AI modifications, particularly those applied without user awareness, can distort perception and recollection. 

\subsection{Positive Use Cases of AI-Generated False Memories}
While the potential for negative impact of AI-generated false memories is significant, it is equally important to explore the positive applications of this technology. When used ethically and under controlled conditions, AI-generated false memories could offer substantial benefits in various fields, particularly in mental health and personal development.

\subsubsection{Therapeutic Memory Reframing}
AI offers immense potential for therapeutic memory reframing~\cite{xiao2024healme, jiang2024generic, lee2019service, mattila2001seeing}, a technique that could revolutionize the treatment of various psychological disorders, particularly those related to trauma and anxiety. In this application, AI might assist in altering distressing memories for patients undergoing psychological treatment, providing a powerful tool for mental health professionals~\cite{xiao2024healme,  jiang2024generic, lee2019service, mattila2001seeing}. By subtly modifying or enhancing specific elements of a photo or video associated with a traumatic event, as shown in Figure~\ref{fig:false-memory-example}, AI could help reduce the emotional intensity tied to these memories. For example, an AI-altered image could remove or blur triggering elements from a scene, allowing individuals to reframe their traumatic experiences in a less distressing way. This technique might be particularly beneficial for patients suffering from Post-Traumatic Stress Disorder (PTSD), phobias, or other anxiety disorders.

The process of memory reframing through AI could involve creating a series of gradually altered images or videos, each one slightly less distressing than the last. As patients work through this sequence with their therapist, they could potentially build new, less emotionally charged associations with the traumatic memory. This approach aligns with existing therapeutic techniques, such as exposure therapy and cognitive restructuring, but offers a more controlled and customizable experience.

It is crucial to note that such applications would require strict ethical guidelines and should only be implemented under the supervision of qualified mental health professionals. The goal would be to aid in the healing process, not to erase or completely falsify memories.

\subsubsection{Enhancing Self-Esteem}
Another promising application of AI-generated false memories lies in the realm of self-esteem enhancement and personal development. AI-edited images of personal achievements or past performances can positively influence self-perception and boost confidence. HCI researchers have already explored the use of AR filters to enhance confidence in public speaking \cite{leong2023picture} and to foster creativity during brainstorming sessions\cite{leong2021exploring}.

With AI, subtly enhancing an image of someone delivering a public speech - perhaps by improving their posture or making their gestures appear more confident, as shown in figure~\ref{fig:false-memory-example}, could reinforce the memory of that success. When an individual recalls this enhanced version of events, it could potentially boost their confidence in public speaking scenarios. 

Similarly, AI could be used to create or enhance visualizations of future success. By generating realistic images of an individual achieving their goals, AI could help reinforce positive self-belief and motivation. This application aligns with visualization techniques already used in sports psychology and personal development coaching~\cite{Weinberg2008, Cumming2013, Renner2019}. It is important to emphasize that the aim here is not to create entirely fabricated memories, but rather to enhance existing positive memories or create vivid, motivating visualizations of future success. The ethical use of this technology would involve full disclosure to the individual and should be implemented as part of a broader personal development or therapeutic program.



\subsection{Ethical Considerations}
The use of AI-generated content to influence human memory raises important ethical concerns that must be carefully considered. Key issues include informed consent and transparency, as individuals should be aware when viewing AI-altered content that may impact their memories or perceptions. The potential for manipulation and misinformation through AI-generated false memories poses risks in various domains, from politics to personal relationships, necessitating robust safeguards against intentional misuse. Privacy and data protection are crucial, both in terms of the datasets used to train AI systems and the potential for highly personalized false memory generation. The psychological impact of AI-induced false memories, particularly in therapeutic contexts, requires careful navigation. Ensuring equitable access to both protective measures and potential benefits is important to prevent uneven distribution of risks and advantages. Researchers, developers, and practitioners have an ethical obligation to consider the consequences of their work and implement safeguards. The need for appropriate regulatory frameworks to balance innovation with protection against harm is evident, as is the importance of considering the long-term societal impact on trust, shared reality, and collective memory. Addressing these ethical concerns requires clear guidelines, robust labeling systems, ongoing interdisciplinary research, enhanced media literacy education, ethical review processes, and collaboration between AI developers, ethicists, policymakers, and cognitive scientists to ensure responsible development and use of this powerful technology.

\subsection{Limitations and Future Research}
This study provides valuable insights into the impact of AI-ge3nerated and AI-edited media on memory formation, but it is not without limitations. These limitations, along with the study's findings, point to several promising directions for future research.

One primary limitation of this study is its reliance on relatively short-term exposure to AI-altered media. Participants were exposed to the manipulated images and videos for a limited time, which may not fully reflect real-world scenarios where individuals are repeatedly exposed to such content over extended periods. Moreover, as demonstrated in Loftus' research, false memory implantation often involves gradual and repeated suggestion rather than a single exposure. However, our study reveals that even brief exposure to these types of stimuli induces a significant false memory effect. Previous studies \cite{chan2024conversational, loftus1995formation, payne1996memory, Frost2000} suggest this effect would likely intensify with increased exposure duration or frequency. Our findings lay crucial groundwork for further investigation, particularly into the cumulative effects of long-term, incremental exposure to AI-generated media on memory formation and distortion. Longitudinal studies could provide valuable insights into how false memories evolve and potentially become entrenched over time with repeated exposure.

Another limitation is the constrained nature of our sample set, both in demographics and dataset size. Demographically, the study was confined to U.S. participants aged 18-100, despite efforts to ensure diversity within these parameters. In terms of stimuli, our study employed only 24 images, potentially limiting the representativeness of AI-generated content. Future research should address these constraints by expanding to a more globally diverse sample and utilizing a larger, more varied image set. This expansion would enhance our understanding of how cultural factors, life experiences, and diverse range of stimuli may influence false memory susceptibility to AI-induced false memories.


The study focused primarily on visual media (images and videos), but AI is capable of generating and manipulating other forms of content, including audio and text. Future research should investigate the impact of AI-edited auditory cues (such as fabricated sound environments) and textual content (like AI-edit personal narratives) on false memory formation. This could provide a more holistic understanding of how different modalities of AI-generated content affect human cognition.


Additionally, a follow-up study should test what difference it makes whether it is the person themself or some other entity that does the editing. This exploration of agency in AI-assisted editing could reveal important insights into the mechanisms of false memory formation and the role of perceived authorship in memory implantation.

While our study examined the formation of false memories, it did not extensively explore the persistence of these memories over time or their resistance to correction. Future research should investigate the longevity of AI-induced false memories and develop effective strategies for correcting or mitigating them once they have formed. Further, our experiment was conducted in a controlled setting, which may not fully capture the complexity of real-world scenarios where individuals encounter AI-generated content. Future studies could employ more naturalistic designs, such as field experiments or ecological momentary assessments, to examine how AI-altered media affects memory in everyday life. This could include studying the impact of AI-generated content on social media platforms or in news consumption.

The ethical implications of AI-generated false memories warrant further investigation. Future research should explore the long-term psychological and social consequences of living in an environment where memories can be easily manipulated by AI. This could include studies of the impact on personal identity, social relationships, and societal trust.
As AI technology continues to advance, future studies should also examine the effectiveness of various countermeasures against AI-induced false memories. This could include evaluating the impact of media literacy training, developing more sophisticated AI detection tools, or exploring the potential of using AI itself to identify and flag potentially misleading content.

\section{Conclusion}
This research demonstrates the significant impact of AI-edited media on human memory distortion. The results reveal that exposure to AI-altered images substantially increases the likelihood of false memory formation, with participants exposed to such content exhibiting a markedly higher propensity to report inaccurate recollections compared to those who viewed unedited control images. This effect was even more pronounced when participants were presented with AI-generated videos based on AI-edited images, suggesting that dynamic AI media significantly amplifies memory distortion. Perhaps most concerning, participants reported high levels of confidence in their false memories, particularly in the AI-generated video conditions.

These findings have far-reaching implications across various domains, including legal proceedings, political discourse, and misinformation.  However, the study also highlights potential positive applications of this technology, particularly in therapeutic contexts. AI-generated content could be used to reframe traumatic memories, or enhance self-esteem when applied ethically and under professional supervision. These findings underscore the need for a balanced approach to AI development that maximizes benefits while mitigating risks.

Moving forward, this research calls for increased awareness and stricter regulations regarding the use of AI in media creation and dissemination.  Future research should focus on developing more effective interventions to mitigate the risk of AI-induced false memories, including improved content labeling systems and public education campaigns. Additionally, interdisciplinary collaboration between AI researchers, cognitive scientists, ethicists, and policymakers will be crucial in addressing the complex challenges posed by AI's influence on human memory and perception. Ultimately, this study serves as a compelling foundation for better understanding and navigating the intricate relationships between AI and human cognition, both now and in the future.

\bibliographystyle{ACM-Reference-Format}
\bibliography{sample-base}

\pagebreak
\section{Supplementary Materials}

\subsection{False Memories Questionnaire}
Please answer the following questions regarding the \textbf{original images}:

\begin{enumerate}
    \item Did you remember seeing the bride smiling in the original picture?
    \item Did you remember seeing the sun in the original picture?
    \item Did you remember seeing this woman with blond hair?
    \item Did you remember seeing a sign on the street?
    \item Did you remember seeing the woman wearing any headwear?
    \item Did you remember seeing a flag in the picture?
    \item Did you remember seeing one of the boys using a smartphone?
    \item Did you remember seeing a boat in the picture?
    \item Did you remember seeing two black guys running in the picture?
    \item Did you remember seeing military presence in the picture?
    \item Did you remember seeing a military bodyguard in the picture?
    \item Did you remember seeing a military vehicle in the picture?
    \item Did you remember seeing only one former US president in the picture?
    \item Did you remember seeing this guy in a mask?
    \item Did you remember seeing guys in hazmat suits walking with a suitcase?
    \item Did you remember seeing coffee mugs on the table?
    \item Did you remember seeing that all the land was covered by snow (no ocean water)?
    \item Did you remember seeing a woman in this group shot?
    \item Did you remember seeing the guy wearing a military uniform?
    \item Did you remember seeing a city in the picture?
    \item Did you remember seeing an adult in the picture?
    \item Did you remember seeing a woman in the picture?
    \item Did you remember seeing a table in the middle?
    \item Did you remember seeing a small bicycle in front of the boys?
\end{enumerate}

How confidence are you with the answer?
\begin{itemize}
    \item 1-Extremely lacking confidence
    \item 2-Very lacking confidence
    \item 3-Somewhat lacking confidence
    \item 4-Neutral
    \item 5-Somewhat confident
    \item 6-Very confident
    \item 7-Extremely confident
\end{itemize}

\subsection{AI filter familiarity}
\begin{itemize}
    \item What is your level of familiarity with using image filter technologies or AI filter technologies? (1-Not familiar at all, 7-Very Familiar)
\end{itemize}

\subsection{Frequency of forgetting}
Participants self-reported memory problems, using a scale ranging from 1 (major problems) to 7 (no problems), as adapted from \cite{zelinski1990memory}.
\begin{itemize}
    \item How would you rate your memory in terms of the kinds of problems that you have? (1= Major problems to 7= No problems)
\end{itemize}

\subsection{Memory Efficacy}
A self-report measure of memory ability (Self-Efficacy Level) taken from a subset of \cite{berry1989reliability}.
\begin{enumerate}
    \item If someone showed me the pictures of 16 common everyday objects‚ I could look at the pictures once and remember the names of 2 of the objects.
    \item If someone showed me the photographs of 10 people and told me their names once‚ I could identify 2 persons by name if I saw the pictures again a few minutes later.
\end{enumerate}
\begin{itemize}
    \item 1-Strongly disagree
    \item 2-Disagree
    \item 3-Somewhat disagree
    \item 4-Neither agree nor disagree
    \item 5-Somewhat agree
    \item 6-Agree
    \item 7-Strongly agree
\end{itemize}

\subsection{Skepticism}
The scale to measure naive skepticism in the adult population is taken from \cite{ferrer2024naive}.
\begin{enumerate}
    \item The official media provides false information
    \item I distrust the information provided by government authorities
    \item The World Health Organization (WHO) hides its true interests
    \item The world press manipulates information
    \item Social networks call those who tell uncomfortable truths crazy
    \item The rich manipulate press
    \item International organizations only deliver information that benefits them
\end{enumerate}
\begin{itemize}
    \item 1-Strongly disagree
    \item 2-Disagree
    \item 3-Somewhat disagree
    \item 4-Neither agree nor disagree
    \item 5-Somewhat agree
    \item 6-Agree
    \item 7-Strongly agree
\end{itemize}

\end{document}